\UseRawInputEncoding
\documentclass[iop, revtex4]{emulateapj}
\usepackage{longtable,lscape}
\usepackage[breaklinks]{hyperref}
\usepackage{natbib}
 
\shortauthors{Galliher et al.}

\begin{document}

\title{Evryscope-South survey of upper- and pre-main sequence solar neighborhood stars}

\author{Nathan W. Galliher\altaffilmark{1}, Jeffrey K. Ratzloff\altaffilmark{1}, Henry Corbett\altaffilmark{1}, Nicholas M. Law\altaffilmark{1}, Ward S. Howard\altaffilmark{1}, Amy L. Glazier\altaffilmark{1}, Alan Vasquez Soto\altaffilmark{1},
Ramses Gonzalez\altaffilmark{1}} 
\altaffiltext{1}{Department of Physics and Astronomy, University of North Carolina at Chapel Hill, Chapel Hill, NC 27599-3255, USA}

\begin{abstract}
Using photometric data collected by Evryscope-South, we search for nearby young variable systems on the upper-main sequence (UMS) and pre-main sequence (PMS). The Evryscopes are all-sky high-cadence telescope arrays operating in the Northern and Southern hemispheres. We base our search on a Gaia-selected catalog of young neighborhood upper- and pre-main sequence stars which were chosen through both astrometric and photometric criteria. We analyze 44,971 Evryscope-South light curves in search of variability. We recover 615 variables, with 378 previously known, and 237 new discoveries including 84 young eclipsing binary (EB) candidates. We discover a new highly eccentric binary system and recover a further four previously known systems, with periods ranging from 299 to 674 hr. We find 158 long-period ($>$50 hr) candidate EB systems, 9 from the PMS and 149 from the UMS, which will allow constraints on the mass-radius-age relation. These long-period EBs include a 179.3 hr PMS system and a 867.8 hr system from the UMS. For PMS variable candidates we estimate system ages, which range from 1 to 23 Myr for non-EBs and from 2 to 17 Myr for EBs.  Other non-EB discoveries that show intrinsic variability will allow relationships between stellar rotation rates, ages, activity, and mass to be characterized. 
\end{abstract}

\section{Introduction} \label{section_intro}

Current stellar models struggle to correctly match observed colors, luminosities, initial mass functions, and age-rotation-activity relations across a range of masses (\citealt{david2019age}). Eclipsing binary (EB) systems serve as the calibration benchmarks on which most of these evolutionary models are built (e.g., \citealt{meng_2014}). Pre-main sequence (PMS) stars are presently particularly under-characterized, due to a paucity of systems; model predictions for these systems have large errors on fundamental qualities such as radius and luminosity (\citealt{hillenbrand_2004}). Evaluating these models depends on comparison with fundamentally determined parameters from known systems, which can be obtained by studying long-period PMS EBs. In general, long-period EBs from any part of the H-R diagram allow for determination of stellar parameters, such as radii and mass for both companions, independently (\citealt{maxted_2018}, \citealt{pavlovski_2010}). 

Young stars also exhibit rotation-induced variability, which are powerful tools for calibrating gyrochronology relations. These relations only require stars with a known rotation period and mass (\citealt{angus_gyro}). Further relationships between stellar rotation rates, activity, and age depend on long-term monitoring of variable stars, across a range of both mass and age (\citealt{coveyage}). Young intrinsically variable stars with observable rotation periods also typically show stellar activity in the form of flares. Active stars exhibiting sinusoidal variations, typically arising from stellar spots paired with star rotation, allow complex relationships between rotation periods, starspot coverage, and stellar activity to be studied (e.g., \citealt{newton_2016}, \citealt{newton_2018}, \citealt{mondrik_2018}, \citealt{howard_2019}).

Ground-based surveys, such as the Palomar Transient Factory (\citealt{law_2009}), ASAS-SN (\citealt{asas_2017}), KELT (\citealt{kelt}), ATLAS (\citealt{atlas}), OGLE (\citealt{ogle_2008}, \citealt{ogle_2014}), HAT/HAT-South (\citealt{hat_2018}, \citealt{hat_2019}), ZTF (\citealt{ztf_2019}), and many more, have proven successful at finding large numbers of variable stars and systems; e.g., \cite{2018MNRAS.477.3145J}, \cite{2019MNRAS.486.1907J}, \cite{2019MNRAS.485..961J}, \cite{jayasinghe2020asas}, \cite{pawlak2019asas}. These surveys typically follow one of two observing strategies: observing the entire sky on day timescales, or reaching higher cadence observations by monitoring specific areas of the sky. 

In this work, we perform a survey looking for young star variability by utilizing the Evryscope-South. The Evryscopes are all-sky, high-cadence telescopes based in the Northern and Southern hemispheres. Each telescope monitors the full sky using 24 6.1 cm aperture telescopes on a shared mount, covering 30 million targets at 2 minute cadence (\citealt{Law_2015}; \citealt{Ratzloff_2019a}). For this survey we utilize the Evyscope-South database, taking advantage of its high-cadence all-sky light curves. The high-cadence observations allow us to detect magnitude variations on short timescales, while multi-year all-sky coverage allows for longer-period magnitude changes to be well sampled. 

This survey joins other Evryscope variability searches, including: A southern sky survey of hot subdwarfs \citep{ratzloff2019hot}; a southern pole survey for high-amplitude variables \citep{ratzloff2019variables}, a flare survey of cool stars in the southern sky \citep{howard_2019}; a survey of cool stars for rotational variability \citep{howard_2019b}; and a survey for super flare occurrences of TRAPPIST-1 \citep{glazier_2019}.

Our data set of 44,971 candidate light curves is based on the \cite{zari_2018} catalog of upper main sequence (UMS) and PMS sources. \cite{zari_2018} uses a series of magnitude, color, and proper motion cuts, paired with extinction and reddening corrections, to develop these catalogs of young stars. In this survey, we recover a total of 615 variable objects, 378 previously known and 237 new discoveries. The new discoveries contain 84 previously unknown candidate EB systems and 153 new stars showing periodic variability. We discovered 5 new PMS EB candidates and an eccentric binary system. For each of our discoveries we estimate mass and provide galactic distances. For PMS systems we also estimate ages and discuss the age-mass distribution. We also highlight a couple examples of stellar activity events.

The paper is presented as follows: Section \ref{section_observations} describes the Evryscope photometry and light curve generation processes as well as details on the target selection process, the algorithms used to detect variability, and fitting parameters. Section \ref{section_discoveries} details the discoveries of EBs, non-EBs, and eccentric EBs. In Section \ref{section_discussion} we draw conclusions and perform analysis on discoveries found in the previous sections, and we also discuss age, mass, and period distributions for our recoveries. Finally, in Section \ref{section_summary} we summarize our results from the survey.

\section{Observations and Variability Search} \label{section_observations}
\subsection{Evryscope Photometry} \label{subsection_photometry}
All EB system and variable star discoveries presented in this survey are the results of Evryscope-South photometric observations taken from 2016 January to 2018 June. The Evryscope-South is a 22-camera array mounted in a 6 ft-diameter hemisphere, allowing an 8150 sq. deg. field of view to be observed per exposure (\citealt{Law_2015}, \citealt{Ratzloff_2019a}). The Evryscope-South takes images in Sloan-\textit{g'} with two minute exposures, giving on average 32,600 epochs of data per target (\citealt{Ratzloff_2019a}). To achieve this all-sky high-cadence coverage the instrument tracks the sky for a two hour period, taking two minute exposures, before ``ratcheting" back and beginning its track of the next sky region (\citealt{Ratzloff_2019a}).  The telescope is at CTIO in Chile and has observed continuously since 2015, providing a database with 16 million sources  and each source with tens of thousands of epochs. 

Here we briefly describe the calibration of images, image reduction, and light curve construction processes. More details on these processes can be found in \cite{Ratzloff_2019a}. Raw images are filtered with a quality check, calibrated with master flats and master darks, and have large-scale backgrounds removed using the custom Evryscope pipeline. Forced photometry is performed using APASS-DR9 (\citealt{Henden_2015}) as our master reference catalog. Aperture photometry is performed on all sources using multiple aperture sizes; the final aperture for each source is chosen to minimize light curve scatter. 

\subsection{Target Selection} \label{subsection_target_selection}
The \cite{zari_2018} catalogs of UMS and PMS stars closer than 500 pc provides the foundation for this survey. We search 44,971 targets in the Evryscope-South database for variability, out of the 56,238 contained in the \cite{zari_2018} catalogs. To select the UMS stars \cite{zari_2018} used the Gaia (\citealt{Gaia_2016}; \citealt{Gaia_2018b}; \citealt{lindegren_2018}) archive and performed magnitude and color cuts ($M_G \leq $4.4 mag, and ($G_{BP}-G_{RP}) \leq$ 1.7 mag). Then, extinction and reddening corrections are applied and another set of magnitude and color cuts are made ($M_{G,0} \leq $ 3.5 mag, and ($G_{BP}-G_{RP})_0 \leq$ 0.4 mag). Galactic distance is calculated for each target and used to cut targets farther than 500 pc from the sun, then tangential velocity cuts are performed ($v_{\perp}<40$ km s$^{-1}$) to ensure the bulk of stars are young and are consistent with stars contained in the disk. This processes of iterative cuts leaves a group of stars with spectral types of O, B, and A (\citealt{zari_2018}). 

\begin{figure*}
\centering
\includegraphics[width = .9 \textwidth]{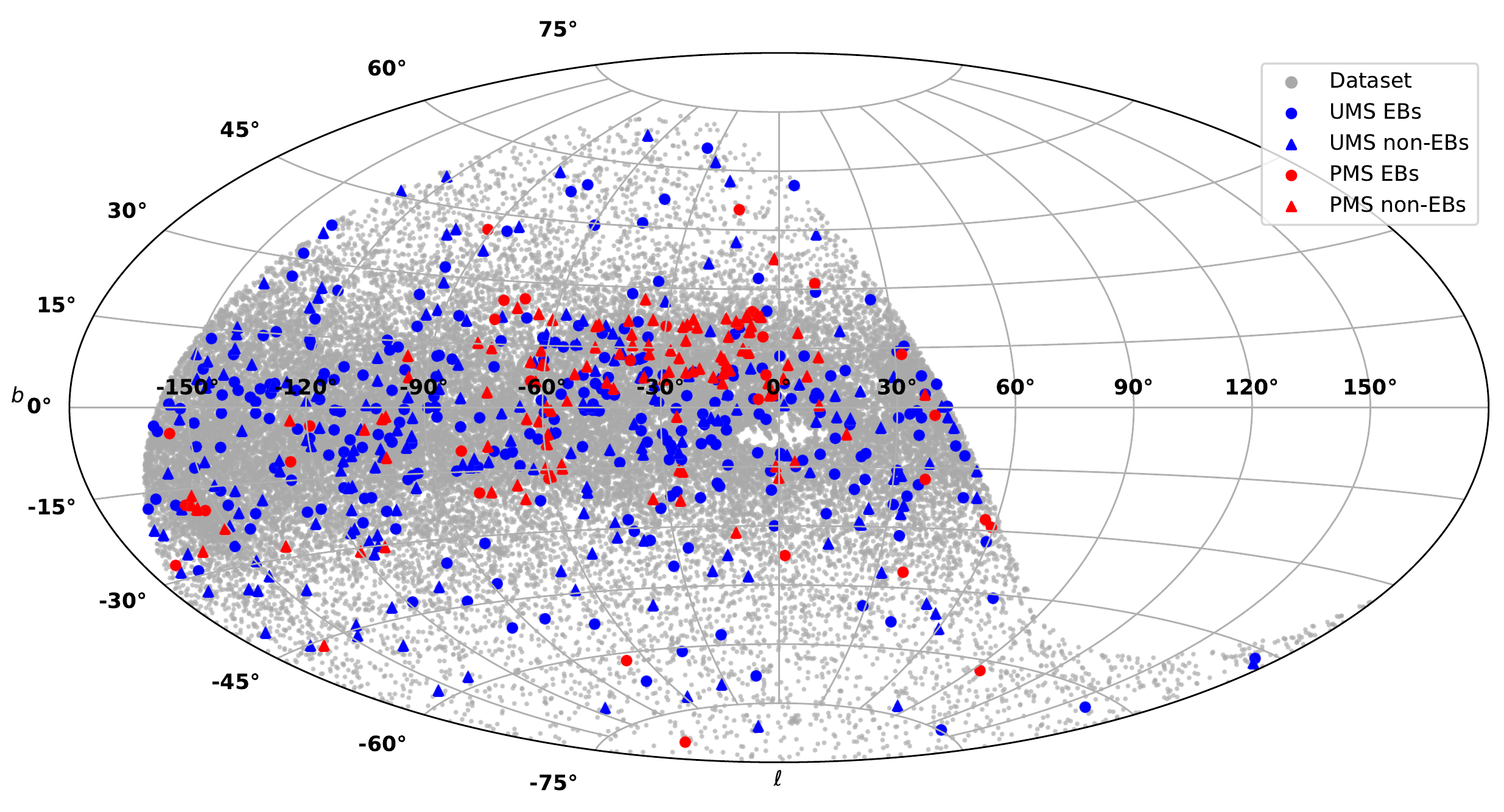}
\caption{An all-sky map of the targets and discoveries from this survey plotted in galactic coordinates. The \citep{zari_2018} data set is shown in gray, with the candidate UMS EBs shown in blue dots, UMS non-EBs shown in blue triangles, PMS EB candidates shown in red dots, and PMS non-EBs shown in red triangles. An empty region is visible in the middle of the Galactic plane, caused by source crowding in the Evryscope input catalog. We see a higher density of PMS discoveries near the galactic center as expected, due to the higher density of PMS sources in that region. The UMS discoveries are less clustered toward the galactic plane.}
\label{all_sky_map}
\end{figure*}

\begin{figure}
\centering
\includegraphics[width = .49 \textwidth]{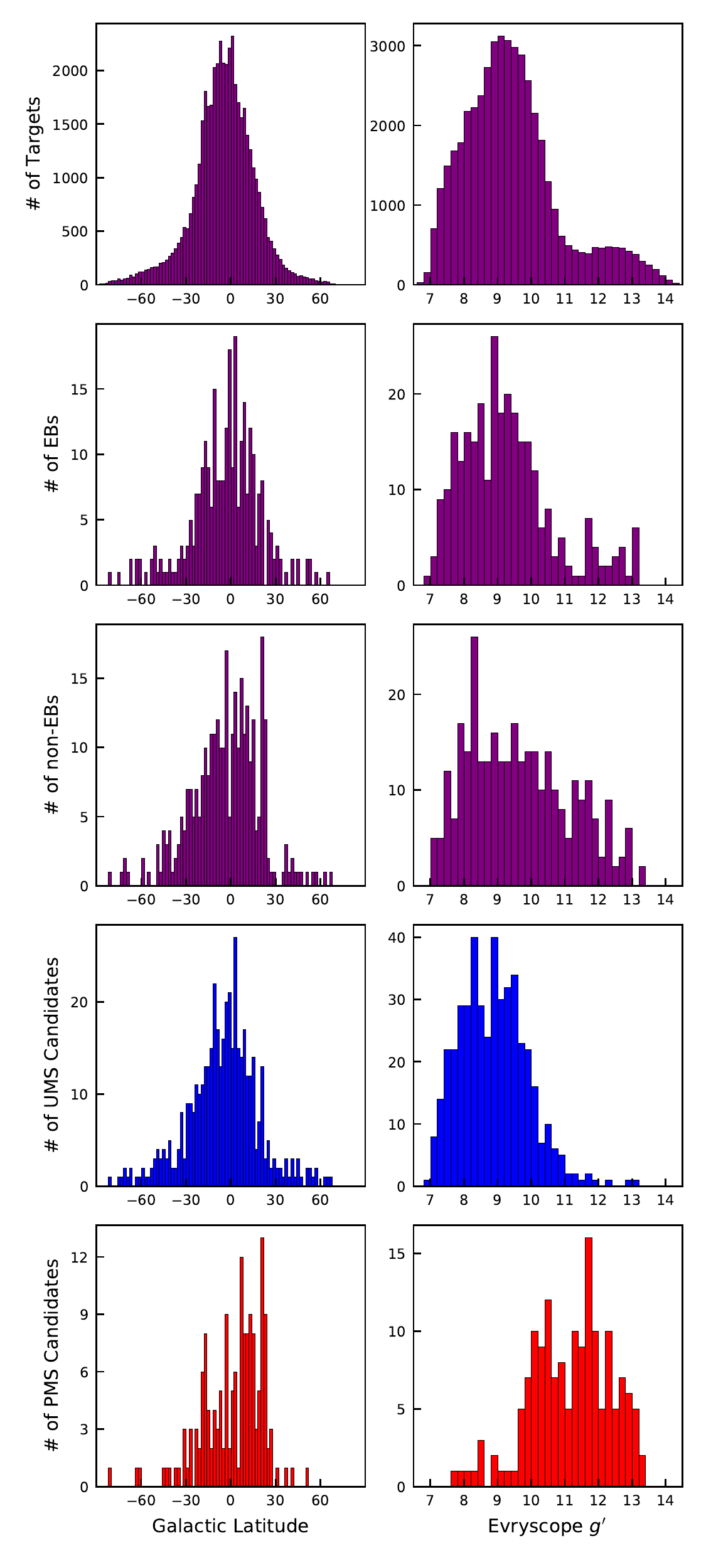}
\caption{Survey targets and discoveries magnitude and galactic latitude histograms; galactic latitude plots have bin sizes of 2$^{\circ}$ and magnitude histograms have bin sizes of .2 mag. The top row displays the galactic latitude (left) and Evryscope-South \textit{g'} magnitude for all targets in the survey; the following two rows show these histograms for all EB candidate discoveries (2nd row from the top) and non-EB candidates (middle row); the final two rows display this information for all discoveries from the UMS (4th row from the top) and PMS (bottom row). The UMS systems tend to be brighter than the PMS systems, while the PMS candidates are slightly more concentrated in the galactic plane than the UMS systems. EB and non-EB discoveries seem to follow similar location and brightness distributions.}
\label{histograms}
\end{figure}

To select the PMS stars from Gaia the same distance and tangential velocity cuts were made as before, to ensure the stars are within 500 pc and contained within the disk. Then, extinction and reddening corrections are applied to the targets, and all stars dimmer than the binary sequence are cut. Stars brighter than $M_{G,0} > $ 4 mag are cut, to exclude stars that are located near the MS turn-off as wells as any giant stars. A second magnitude cut is made so that only stars brighter than the 20 Myr isochrone are included. To remove any remaining MS stars, any source with $A_G > $ .92 mag is discarded, leaving a catalog of PMS stars (\citealt{zari_2018}). An all-sky map of the target coordinates can be found in Figure \ref{all_sky_map} and target distributions of magnitude and galactic latitude are shown in Figure \ref{histograms}.

\subsection{Gaia Cross-Matching and Contamination}
The Evryscope-South has a pixel size of 13.1$\arcsec$ for single images and the Evryscope photometry pipeline is capable of achieving $1\arcsec$-$2\arcsec$ RMS astrometry, with point-spread function (PSF) FWHMs ranging from 0.5 to 5 pixels. The variable aperture size used in the Evryscope pipeline is chosen such that the instantaneous photometric accuracy, averaged over the light curve, is optimized. Compared with the input catalog that is based on Gaia, the Evryscope has extremely large pixels and PSF sizes. This implies the possibility of mismatching or contamination between the input catalog sources and queried Evryscope light curves. In order to understand the level of possible contamination introduced by the large Evryscope pixel size we searched for nearby bright sources around each of our candidate discoveries. We find that 5\% of UMS and 13\% of PMS discoveries have nearby neighbors bright enough to cause possible variability contamination with the intended target source. A similar Evryscope survey, which analyzed 160,000 bright stars near the south pole, found 649 of the targets showed signs of variability, or $\mathtt{\sim}$.4\% \citep{ratzloff2019variables}. This implies the likelihood that our candidate discoveries are contaminated by variable field stars nearby the target stars is low, around .02\% of UMS and .05\% of PMS candidates. A more likely source of contaminants is the input target list, where it is possible the UMS and PMS lists may have contaminants caused by reddening and other selection effects.

\subsection{Detection Methods} \label{subsection_detection_methods}
All 44,971 light curves were searched for variability using conventional period search algorithms and a specially developed ``outlier'' detector. Below we provide the algorithm settings for conventional tools as well as a brief description of the custom outlier tool. Next, we provide details on how these tools are used to select targets for visual inspection, determine the correct period for detections, and vet stars that might show variability due to Evryscope or search algorithm systematics.

\subsubsection{Box Least Squares}
The Box Least Squares (BLS) algorithm (\citealt{kovacs_2002}, \citealt{ofir_2014}) is applied to all targets to obtain a periodogram. For the BLS algorithm we used 25,000 test periods with a range of 3.0-240.0 hours, and with transit fractions ranging from 0.01 to 0.25.

\subsubsection{ Lomb-Scargle}
The Lomb-Scargle (LS) algorithm (\citealt{Lomb_1976}, \citealt{scargle_1982}) is applied to all targets to obtain a periodogram (power spectrum). For the LS algorithm we used a period range of 3.0-720.0 hours. The Lomb-Scargle periodogram applied in this survey is the \textit{LombScargle} Python module from \textit{Astropy} (\citealt{astropy:2013}, \citealt{astropy:2018}). This module automatically selects a period spectrum for the LS periodogram, based on the input period test bounds and epochs, in order to optimize both speed and period recovery.
\\
\subsubsection{Outlier Search Algorithm}
The custom outlier search algorithm is a tool developed in \cite{ratzloff2019hot} and designed to find short, deep transit events. The light curve is normalized in flux space, and the 1$\sigma$ error is computed. Next, data points are selected that are below 3$\sigma$ from the mean value. The number of points below this threshold is compared to a predetermined minimal number, which for this survey is set at 100 to ensure most transits will be well populated. If the number of points below this 3$\sigma$ from the mean value is not met, more points are iteratively included in the selection in steps closer to the mean by .1$\sigma$ until the minimal threshold is met. Once the threshold is met, all points selected by this iterative process are then phased folded at 125,000 periods ranging from 1.0 to 240.0 hours. At each period the standard deviation is calculated in phased time, ignoring the normalized flux values of the outlier points. Power is calculated using these standard deviations of outlier points in phased time. The minimum value from these standard deviations is selected as the best period. Details on this algorithm and its performance can be found in \cite{ratzloff2019hot}. 

\subsubsection{Application of Search Algorithms and Classification}\label{algorithm_application}
\begin{figure}
\centering
\includegraphics[width = .49 \textwidth]{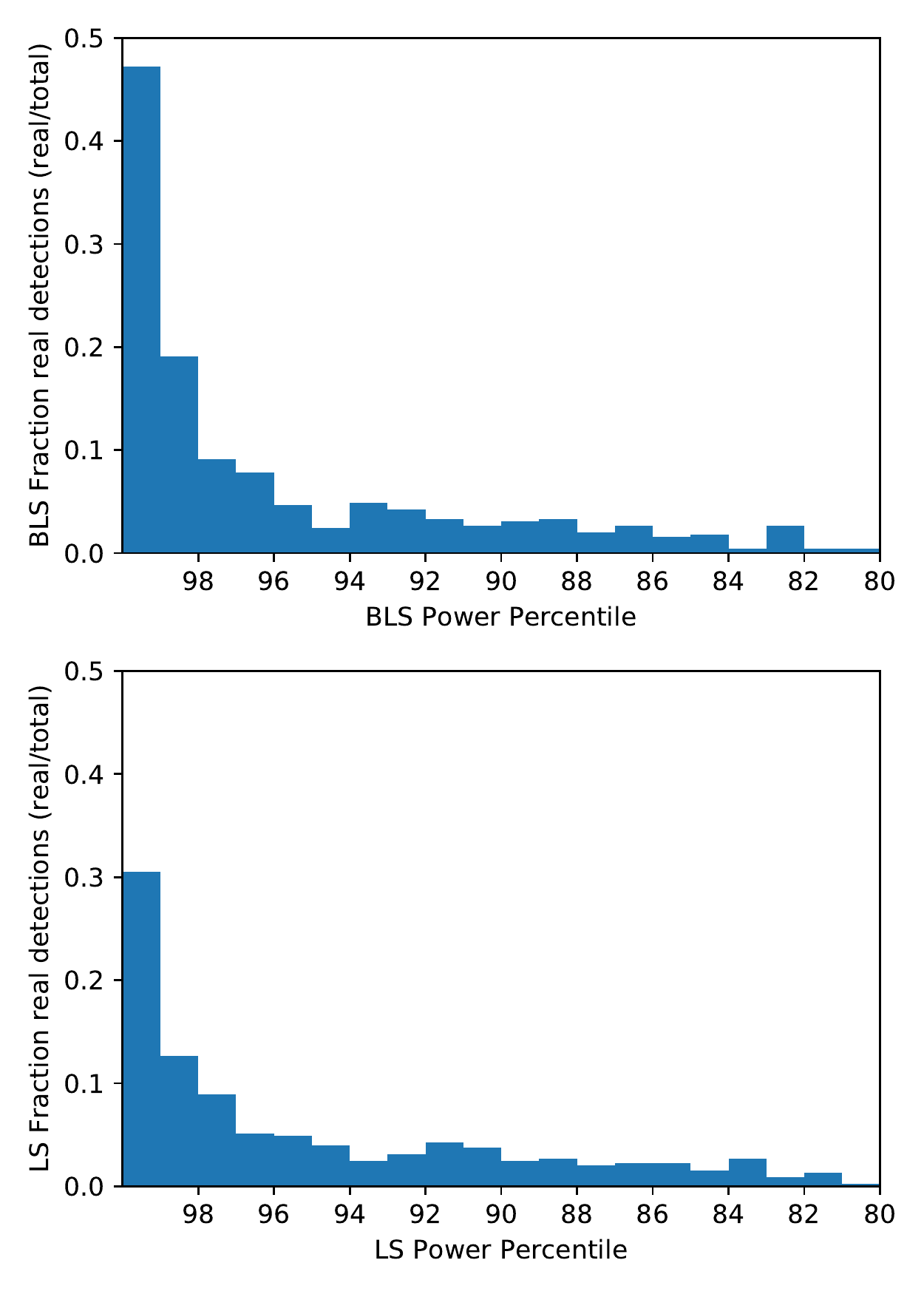}
\caption{BLS (top) and LS (bottom) plots of periodogram power percentile vs the fraction of detections that were real. This shows the likelihood of a detection to be real given the percentile of the periodogram's power. Both have bin sizes of one percentile.}
\label{per_vs_frac_det}
\end{figure}

We first apply all three search algorithms to all 44,971 light curves in our data set, obtaining 3 periodograms per light curve. For each of these periodograms common systematics such as the day/night cycle are filtered out, as described in \cite{ratzloff2019variables}. The light curves with the top 20\% of BLS and/or LS powers are visually inspected for variability. Targets below the 80th percentile are not visually inspected because returns from searching drop off below this detection level (Figure \ref{per_vs_frac_det}). 

Visual inspection of light curves proceeds in 3 stages. The first stage is used to determine if any periodic variability exists in the light curve. The selected targets are all phase folded at the following periods: top 3 BLS powers, top LS power, and minimum outlier standard deviation. Along with the folded light curves, the light curve and the periodogram for each of the search algorithms are visually inspected simultaneously. This allows for comprehensive viewing of all detection signals as well as comparisons between different search algorithm periods. During this process the discoveries of periodicity are primarily made by the BLS or the LS algorithm. The outlier method complements these discoveries by narrowing down the exact period, especially for EB systems. 

The second stage of visual inspection is a secondary vetting to make sure variability found in the first stage is produced by the target star and not a systematic. We query near the target R.A. and decl. for light curves from reference stars that have similar magnitudes, requiring that they be greater than 60$\arcsec$ away to prevent blending. We visually inspect the 3 closest reference stars' light curves for variability at the same period as the target star's period. 

The final stage of visual inspection is used to determine whether the correct period or an alias thereof was initially detected in the first stage of vetting. We fold each target at half and double the initial period. This is useful for many systems that can display double eclipses of similar, but different depths.

During the visual inspection process targets are identified as belonging to one of two groups. Stars showing intrinsic variability in the form of sinusoidal-shaped light curves are labeled as variable (non-EB) stars. Objects showing extrinsic variability, entirely in the form of eclipses, are labeled as EB if they show one eclipse or EB2 if they show two eclipses. Many cases exist where identification can be subjective, such as EBs that are in close orbits; the light curves of these objects can appear to be sinusoidal. In these cases we looked for sharp transitions in the light curves, multiple dips with different depths, and asymmetry in the rise and fall times of variability to identify EBs from variable stars. In order to determine the classification in these situations we start by plotting the brightness dips (possible eclipsing events) separate from the full light curve. This allows a close look at the potential eclipse, in search of a flat bottom or a sharp peak representative of an EB. At this point, we also refer back to the period alias analysis performed earlier to check for possible brightness decreases of different depths or shapes, which might indicate a contact binary system. Next, the full unfolded light curve is inspected to check for systematic errors in the light curve and to look at individual brightness changing events in hopes of better deducing the overall shape of the curve. In these borderline cases with potential EB candidates at short periods, where the eclipses are long enough, and the tidal distortions become large, the light curve effects blend together making classification subjective. This is discussed more in Section \ref{short_period_variables_sec}. In the cases where this process does not make the classification any clearer we classify the object as being a non-EB.

A final stage of identification is performed on the variable stars to identify a small subset that showed variability that was non-sinusoidal. These objects are labeled periodic non-sinusoidal (NS), and the objects showing sinusoidal variations are labeled sinusoidal variables (S). For clarity in this manuscript, when discussing stars showing intrinsic variability of either kind (NS or S) we refer to the systems as non-EBs.

\subsection{Parameter Estimation} \label{subsection_fitting}
\begin{figure*}
\centering
\includegraphics[width = \textwidth]{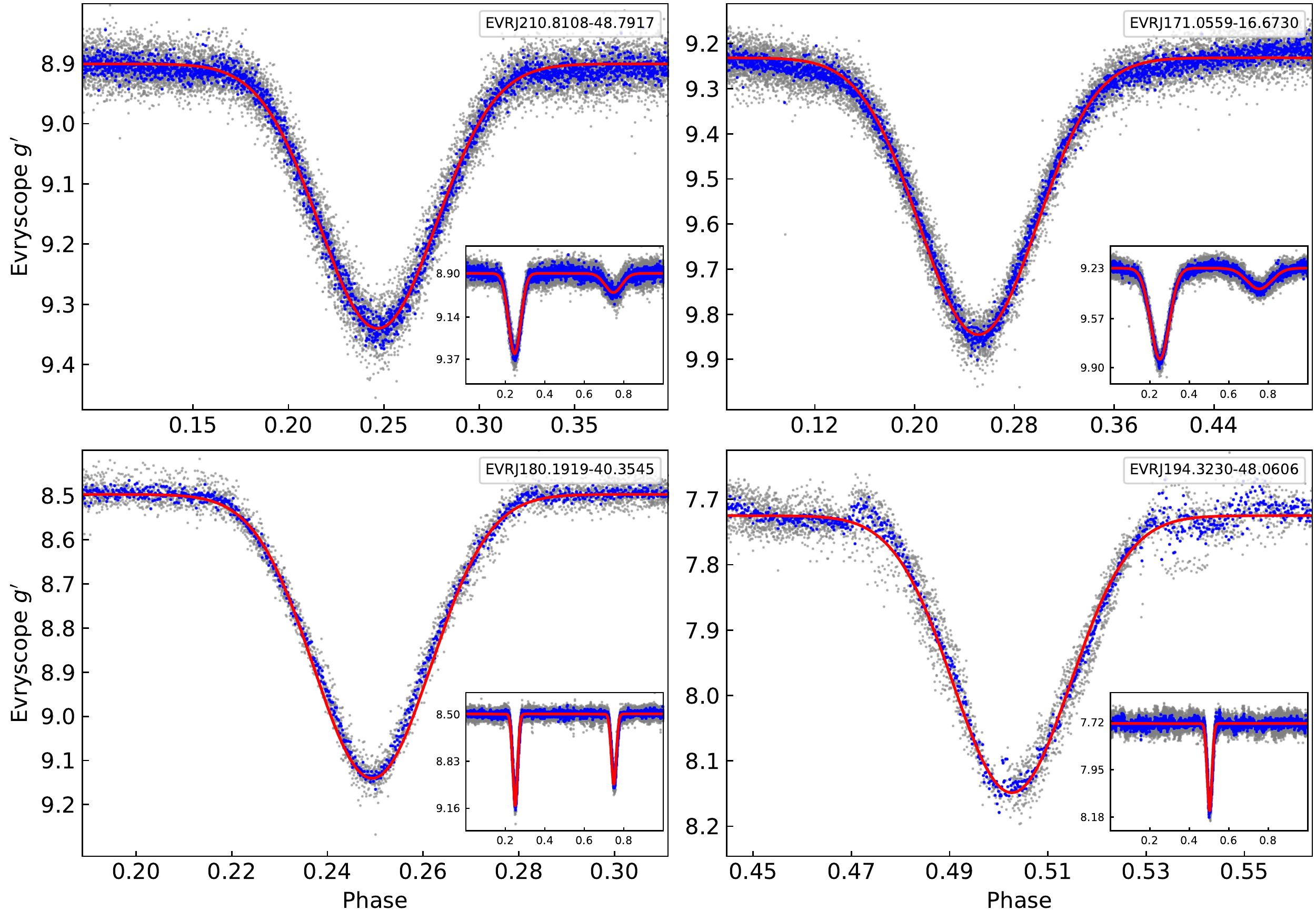}
\caption{Examples of Gaussians fit to four of the EB candidates found in this survey. The large figure in each panel shows a zoomed in plot of each eclipse and the subplot contains the full light curve. Each of these plots show the raw Evryscope data in gray, the binned in phase data in blue with 15 points per bin, and the Gaussian fit in red. }
\label{plot_fitted_transits}
\end{figure*}

After the discovery and vetting stages each target is reprocessed with BLS or LS depending on whether they were EBs or intrinsically variable stars, respectively. The periodogram is calculated in a narrowed range from 95\% to 105\% of the target's period in order to find a more accurate period. We use this period to fit each target with one of three possibilities depending on the variable type: Non-EBs are fit with a sinusoid, EBs with only one eclipse are fit with a single inverted Gaussian peak, and EBs with two eclipses are fit with double inverted Gaussian peaks. We note here that while a Gaussian fit is not a physical model for the EB candidates, these fits do capture many of the desired features. Gaussian fits allow for sufficiently accurate estimates of depths, durations, and zeroth epochs for each of the candidates. A few examples of these fits are displayed in Figure \ref{plot_fitted_transits}. Fits are performed using the scipy \textit{curve\_fit} Python routine (\citealt{scipy}, \textit{scipy.optimize.curve\_fit}).

\subsection{Uncertainty Determination}\label{subsection_error_determination}
\subsubsection{Period and Amplitude Uncertainty}\label{subsection_amp_per_err}
For EB systems we allow cumulative error in phasing over the years-long light curves to be equal to one half the width of the main eclipse. We conservatively assume that past this point the light curve shape would distort and become apparent in the phased light curve. In the same manner, for non-EBs we allow the total cumulative error in the phased light curve to be one quarter of the total period of the system and divide this by the total number of periods in the light curve to estimate the period uncertainty. 

We estimate fit uncertainties using the covariance matrix returned from \textit{curve\_fit}. We fit the respective variable model (Gaussian, double Gaussian, sinusoid), and input the Evryscope magnitude uncertainties. Each fit is visually inspected for accuracy. We set a minimum uncertainty of 0.003 mag, arising from Evryscope systematics. 

\subsubsection{Distance Errors}\label{subsection_dist_err}
\begin{figure}
\centering
\includegraphics[width =.49 \textwidth]{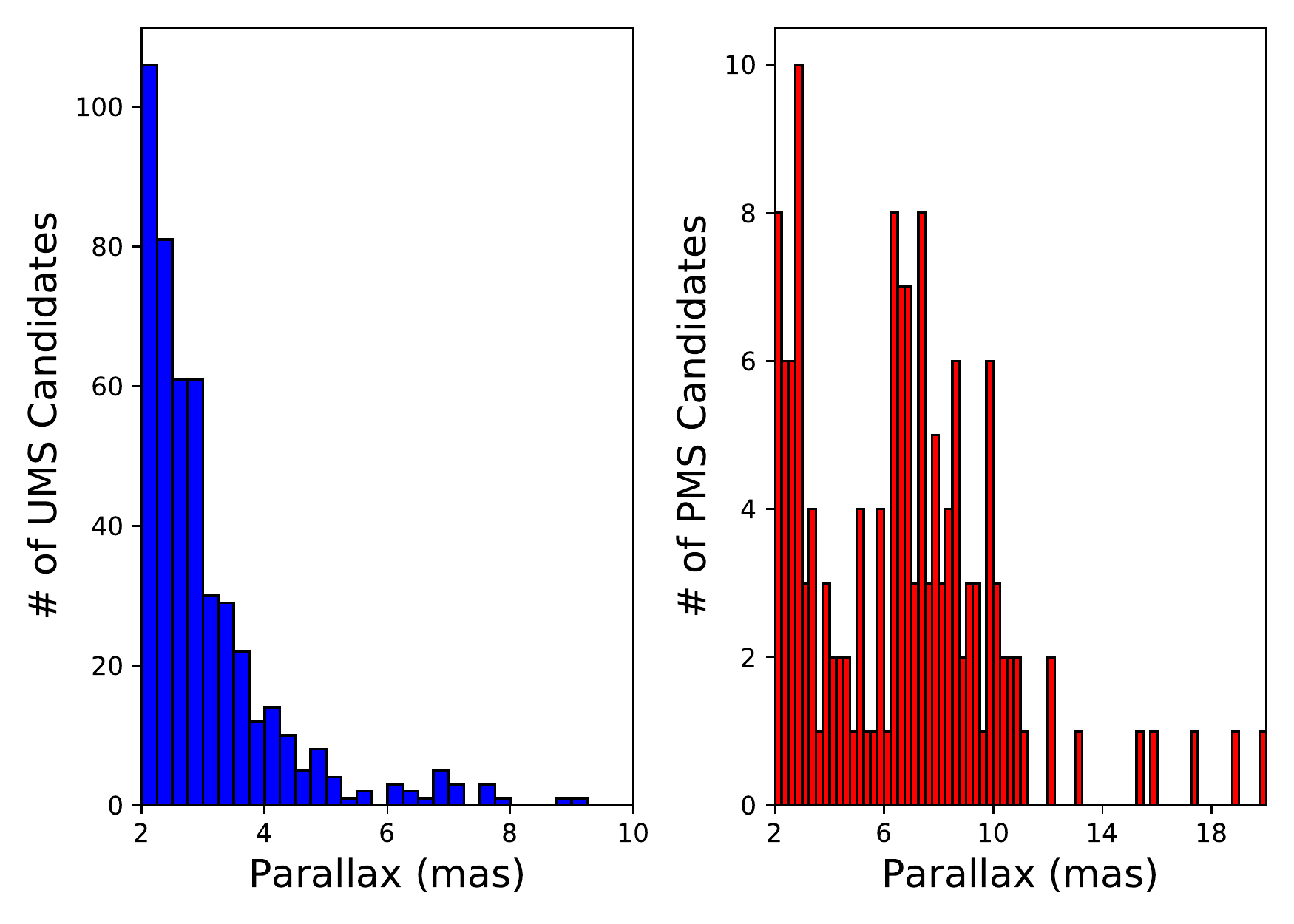}
\caption{Parallax distributions for UMS discoveries (left) and PMS discoveries (right). Both plots have bin sizes of .25 mas. The PMS systems show two clusters, around 2 and 7 mas, while the UMS systems only show one peak at 2 mas which falls off exponentially with parallax. }
\label{parallax_distributions}
\end{figure}

Gaia parallax values and errors, including the Gaia systematic parallax error of $\mathtt{\sim}$.04 mas \citep{lindegren_2018}, are used in a Monte Carlo simulation in order to obtain final galactic distance values and errors. We use 10,000 samples for each target and take the median of the trials to be the distance value and the standard deviation of the trials is taken as the error. The parallax distributions for EB and non-EB candidates is shown in Figure \ref{parallax_distributions}. The UMS systems tend to be closer and we see lower numbers of systems for larger parallaxes, but the PMS systems show two clusters around parallaxes of 3 and 5 mas.

\section{Discoveries} \label{section_discoveries}
\begin{figure*}
\centering
\includegraphics[width =  \textwidth]{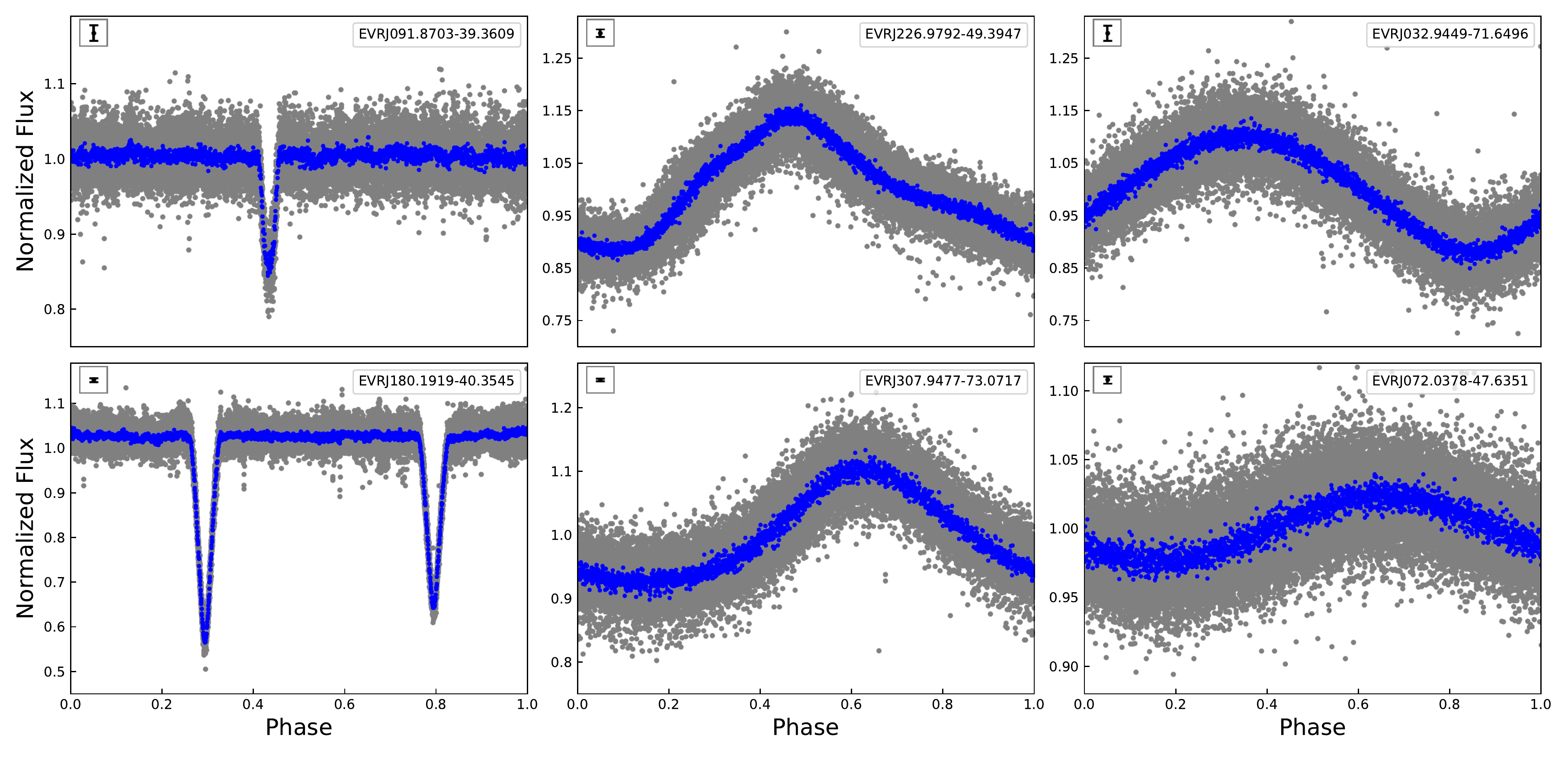}
\caption{Selected Evryscope example light curves showing eclipsing binary with single eclipse [EB] (top left), eclipsing binary with two eclipses [EB2] (bottom left), periodic non-sinusoidal variables [NS] (center), sinusoidal variables [S] (right). The gray points show the raw Evryscope data and the blue points show the phase-binned data with 15 points per bin. A representative error bar showing the average 1-$\sigma$ uncertainty of each raw photometric point is included in the upper left of each plot in a gray box.} 
\label{lc_examples}
\end{figure*}

Using the labels defined in Section \ref{algorithm_application} our search recovered 302 candidate binary systems (EB or EB2), 295 sinusoidally varying objects, and 18 periodic non-sinusoidally varying objects. Figure \ref{lc_examples} shows some randomly selected examples of Evryscope light curves for these classifications. For the purpose of some discussions the classifications of S and NS are grouped as non-EB systems, while EB and EB2 are grouped generally as EBs. We note that some non-EB systems may actually be binary systems that are indistinguishable by only inspecting light curves, while still others may show variability arising from contaminants in the sample such as radial pulsators. An overview of the period distributions is shown in Figure \ref{discoveries_period_histograms}, and Evryscope-South magnitude and galactic latitude distributions are shown in Figure \ref{histograms}. We find that in general the UMS candidates are brighter in the Evryscope-South bandpass than the PMS systems, and the UMS systems are also more spread throughout the galactic plane than the PMS recoveries. Tables for both the UMS and PMS (Tables \ref{ums_table} and \ref{pms_table}, supplementary materials) discoveries are given in the Appendix. An all-sky plot for discoveries from this survey can be found in Figure \ref{all_sky_map}.

\subsection{Eclipsing Binaries} \label{subsection_eclipsing_binaries}
This survey identified 302 candidate EB systems in the southern sky, of which there are 84 new EB systems not previously identified or that had unknown periods. The light curves for all of the new EB candidate systems are shown in Figures \ref{large_lc_plt_first_45} and \ref{large_lc_plt_last_39} (included at the end of the manuscript due to size). These systems have been separated into 2 tables, based on their location in either the UMS or the PMS. Table \ref{pms_eb_table} contains the 5 new candidate PMS EB discoveries, specifically highlighted because these systems are uncommon and very useful for stellar model calibrations. Table \ref{eccentric_eb} contains 5 eccentric binary systems found, to be discussed more in Section \ref{subsection_highly_eccentric_binaries}. Table \ref{ums_table} (supplementary materials) contains the 79 new candidates located in the UMS. These systems are spread throughout the southern sky, with a concentration near the galactic plane. Periods recovered for these EB candidates ranged from a few hours for very close contact binaries to over a month for the longest system. This is a result of both our large input search parameters, searching for periods over both short and long timescales, and at small and large amplitudes. This survey was strong at recovering new discoveries with low amplitudes, with most of our discoveries having primary eclipses smaller than .3 mag. The distribution of period amplitudes can be seen in Figure \ref{discoveries_period_histograms} (bottom), and the distribution of primary transit amplitudes can be seen in Figure \ref{amp_histograms} (bottom).

\tabcolsep=0.09cm
\begin{deluxetable}{llll}
Evryscope ID & Period & Amp. & Sec. Amp. \\
 & (h) & (mag) & (mag) \\
\hline
\endhead
EVRJ095.0357+04.9090* & 142.60$\pm$0.02 & 0.37$\pm$0.01 & 0.35$\pm$0.01\\
EVRJ117.1324-68.2810 & 179.332$\pm$0.009 & 0.483$\pm$0.003 & --- \\
EVRJ175.6080-79.5222 & 91.824$\pm$0.003 & 0.17$\pm$0.02 & 0.11$\pm$0.01 \\
EVRJ260.8893-32.2224 & 15.215$\pm$0.001 & 0.19$\pm$0.02 & 0.19$\pm$0.02 \\
EVRJ359.4872-02.5466 & 55.825$\pm$0.002 & 0.74$\pm$0.05 & 0.50$\pm$0.04 \\
\hline
\\
\caption{List of new PMS eclipsing binary candidates. `Amp.' and `Sec. Amp.' indicate the amplitudes of the primary and secondary eclipses, respectively.}
\label{pms_eb_table}
\end{deluxetable}

\subsection{Non-EB Variable Stars} \label{subsection_variable_stars}
Our survey revealed 153 new discoveries of non-EB variable star candidates that were either previously unknown or had no known period. Of these new discoveries 118 of the stars were from the UMS while the remaining 35 were from the PMS. Out of the 313 non-EB discoveries 18 of them showed variability that was not sinusoidal. These were given the label NS (periodic non-sinusoidal, Tables \ref{ums_table}, \ref{pms_table}, supplementary materials). This subset of systems could have a wide range of star types, ranging from RR Lyraes to extremely close contact binaries. This might suggest our input catalog has some contaminants, as we would not expect to see RR Lyraes in the UMS or PMS data sets. The rest of the non-EB candidates that showed sinusoidal variability were given the label S (Tables \ref{ums_table}, \ref{pms_table}, supplementary materials).  

\begin{figure}
\centering
\includegraphics[width = .49 \textwidth]{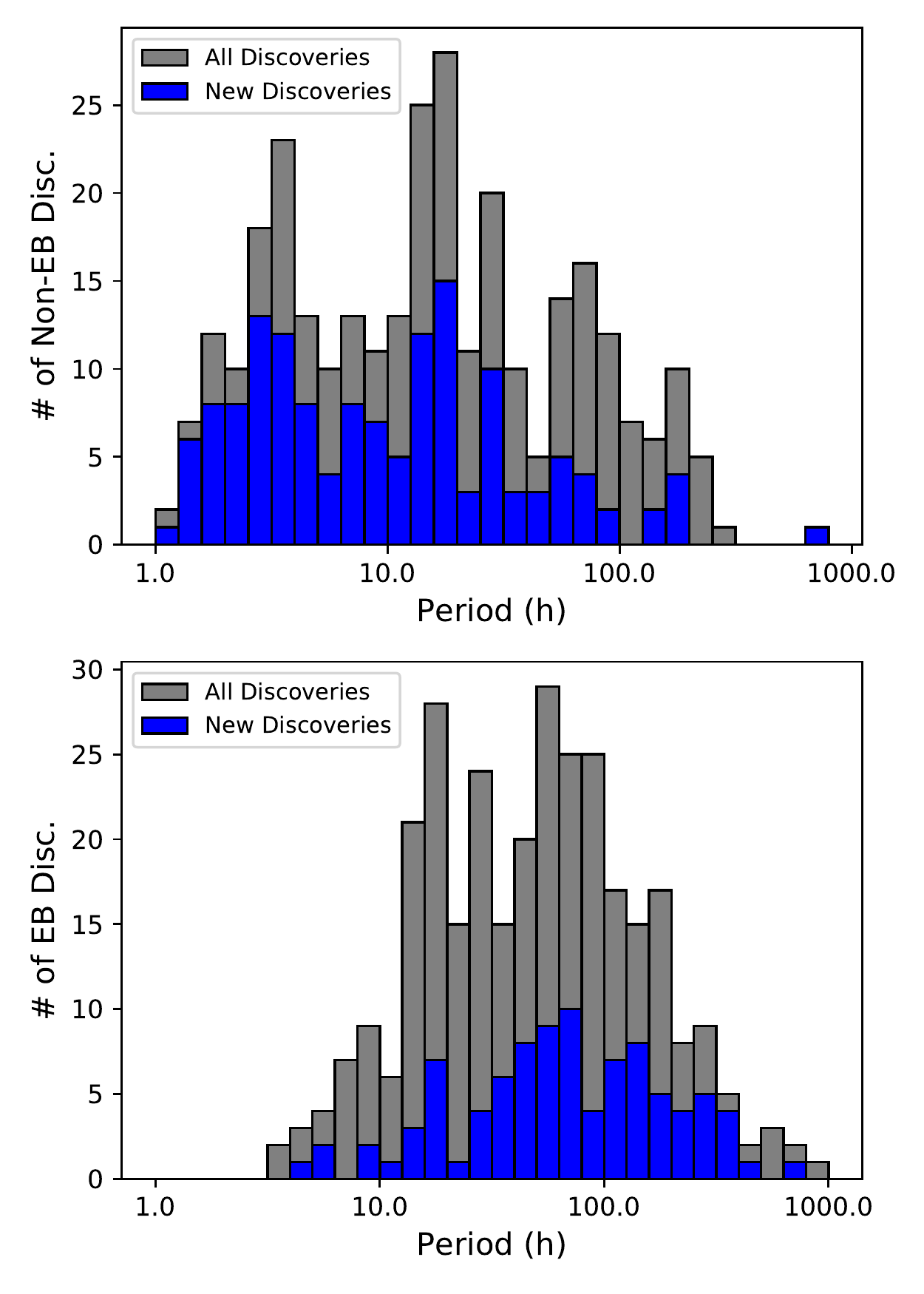}
\caption{Period distributions for variable star (top) and eclipsing binary (bottom) discoveries. The gray bars indicate the number of total targets in that bin, both previously known and unknown. The blue bars represent new discoveries. Both plots have bin sizes of .1 in log(period[h]) space.}
\label{discoveries_period_histograms}
\end{figure}

\begin{figure}
\centering
\includegraphics[width = .49 \textwidth]{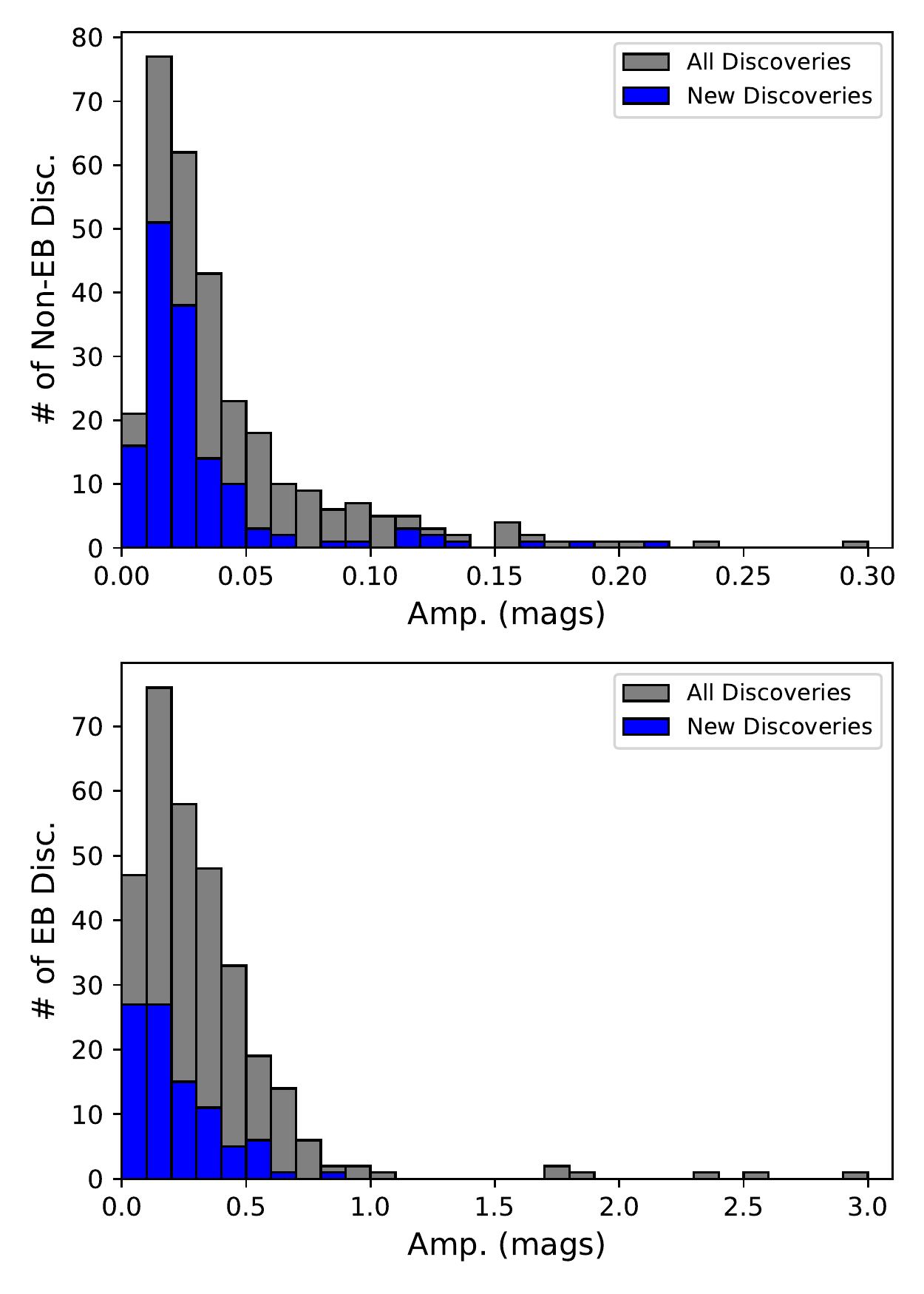}
\caption{Amplitude distributions for non-eclipsing binary (top) and eclipsing binary (bottom) discoveries. The gray bars indicate the number of total targets in that bin, both previously known and unknown. The orange bars represent new discoveries. Both plots have bin sizes of .01 mag.}
\label{amp_histograms}
\end{figure}

All non-EB candidates are fit with a sine wave, and their amplitudes recorded in the data tables (Tables \ref{ums_table}, \ref{pms_table}, supplementary materials). The amplitudes given represent half the peak to trough height of the variation in magnitude. Many of these new discoveries exhibit relatively shallow amplitudes, as seen in Figure \ref{discoveries_period_histograms}. This survey was sensitive to a wide range of period and amplitude variations. We recovered periodic variability ranging from a little more than an hour all the way to periods over a month in duration. The years of high-cadence data contained in the Evryscope database allowed us to achieve sensitivity to very small amplitude variations (on the order of 0.003 mag), by binning the light curves in time. The majority of our non-EB recoveries had amplitudes smaller than .03 mag, while we were still sensitive to variations at much larger amplitudes. The distribution of period amplitudes can be seen in Figure \ref{discoveries_period_histograms} (top), and the distribution of primary transit amplitudes can be seen in Figure \ref{amp_histograms} (top).

\subsection{Eccentric Binaries} \label{subsection_highly_eccentric_binaries}
\begin{figure}
\centering
\includegraphics[width = .49 \textwidth]{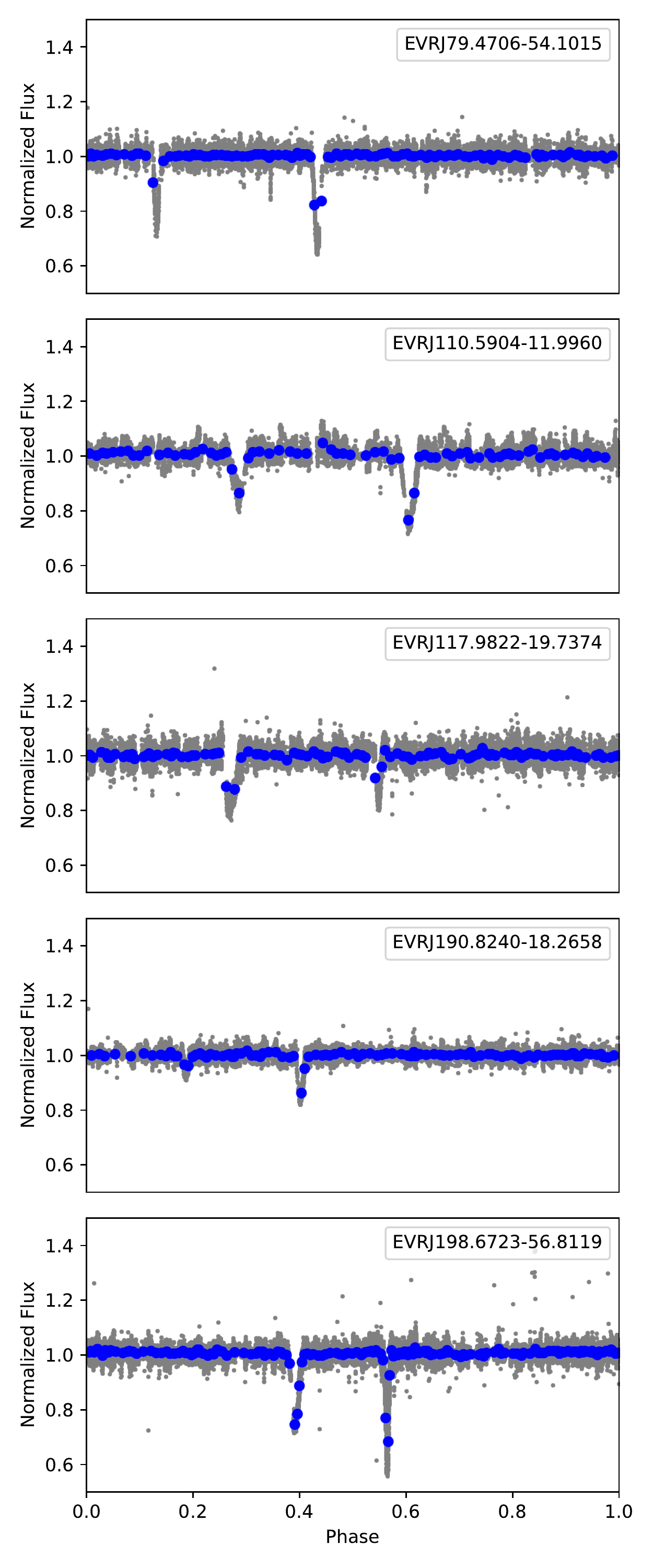}
\caption{Eccentric binary candidates found in the survey. The gray points are the raw Evryscope light curves after filtering for flagged points, while the blue points are binned flux at 200 points per bin.}
\label{eccentric_binaries_combined_plot}
\end{figure}

The five candidate eccentric binaries found in this survey are presented in Table \ref{eccentric_eb}, with light curves for each in Figure \ref{eccentric_binaries_combined_plot}. All of these systems are from the UMS. One of these is a previously unknown EB system, \textit{EVRJ117.9822-19.7374}, with parameters provided in the referenced table.  These systems have periods ranging from 299 to 674 hr, and they have primary eclipse amplitudes between .007 mag and .51 mag. These systems are specifically noted here because some eccentric binaries allow for characterization of the internal composition for each component.

\tabcolsep=0.09cm
\begin{deluxetable}{llll}
\label{long_table}
Evryscope ID & Period & Amp. & Sec. Amp.  \\
 & (h) & (mag) & (mag)  \\
\hline
\endhead
EVRJ079.4706-54.1015 & 627.13$\pm$0.07 & 0.45$\pm$0.01 & 0.33$\pm$0.02  \\
EVRJ110.5904-11.9960 & 312.34$\pm$0.04 & 0.321$\pm$0.008 & 0.175$\pm$0.008\\
\textit{EVRJ117.9822-19.7374}  & 299.31$\pm$0.02 & 0.26$\pm$0.01 & 0.13$\pm$0.01  \\
EVRJ190.8240-18.2658 & 450.72$\pm$0.04 & 0.190$\pm$0.007 & 0.076$\pm$0.005  \\
EVRJ198.6723-56.8119 & 674.62$\pm$0.07 & 0.51$\pm$0.01 & 0.34$\pm$0.01  \\
\hline
\\
\caption{List of eccentric eclipsing binary candidates. `Amp.' indicates the amplitude of the primary eclipses. All of these candidates are UMS systems. Italics indicates previously unknown systems.}
\label{eccentric_eb}
\end{deluxetable}

\section{Stellar Age, Mass, and Activity} \label{section_discussion}
To obtain mass estimates we use PARSEC isochrones (\citealt{bressan_isochrones}) version 1.2S (\citealt{chen_2014}; \citealt{tang_2014}; \citealt{chen_2015}) with $Z = 0.0152$ (solar metallicity) and $A_V = 0$ mag with ages between $10^6$ and $10^8$ yr. We chose to use PARSEC isochrones for consistency with the \cite{zari_2018} input catalog and because these models have been adapted to handle PMS systems. PARSEC uses equations of state, opacities, nuclear reaction chains, neutrino loss, convection, diffusion and many other relevant solar mechanisms to produce models of stellar tracks on the main and pre-main sequence. Using these tracks, we form a KDTree (\citealt{scipy}, \textit{scipy.spatial.KDTree}) using the $BP-RP$ and $M_G$ values from the isochrones. This allows us to query the isochrone models for the nearest neighbor points, in magnitude-color pairs, for each of our targets. For each of the discoveries a mass is estimated by first finding the three closest isochrone color-magnitude pairs to the target color-magnitude pair using the KDTree. The masses are obtained by taking an average of the three closest points masses weighted by their respective distances in the KDTree. 

\begin{figure}
\centering
\includegraphics[width = .51 \textwidth]{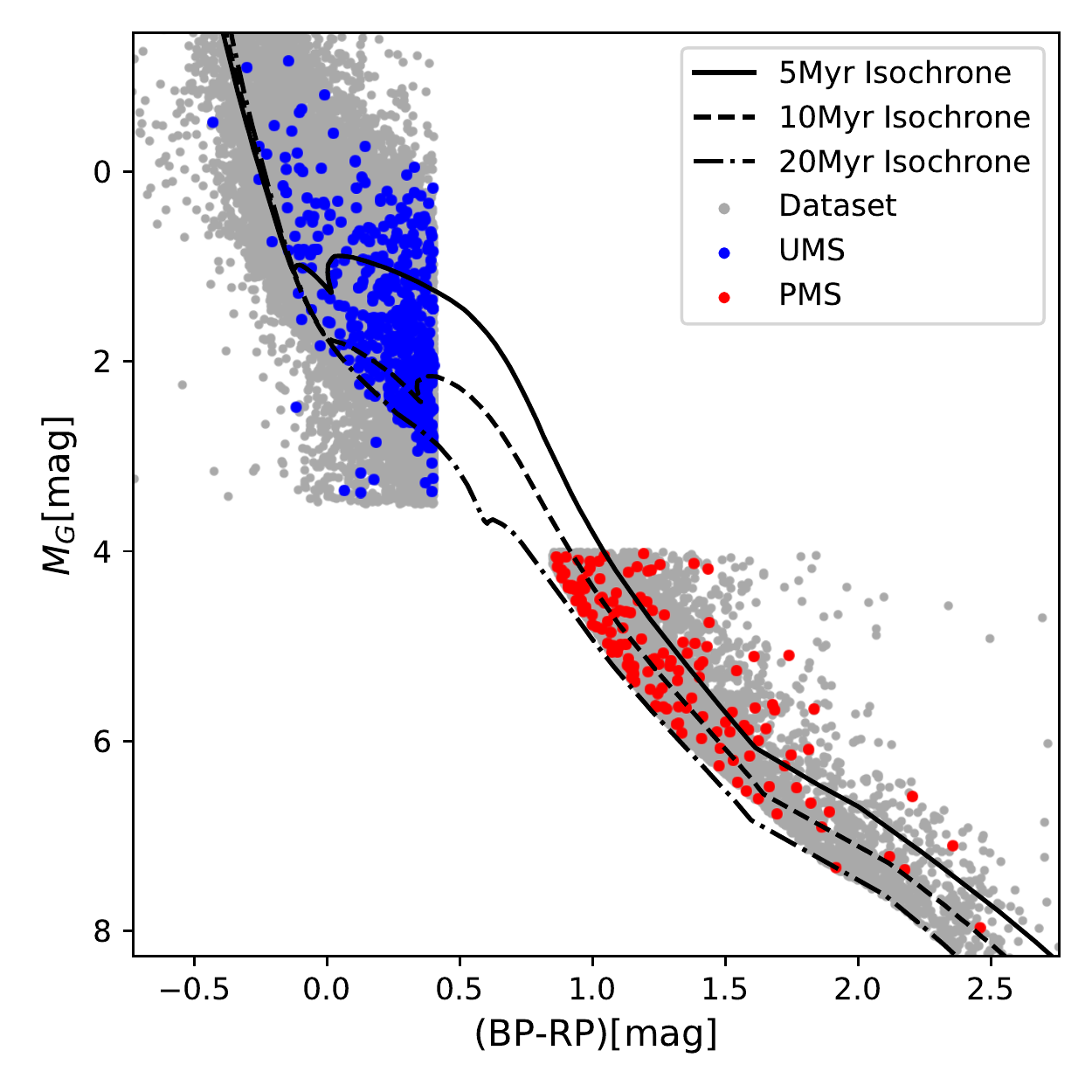}
\caption{Color vs $M_G$ plotted for all targets in this survey. Blue points are UMS systems, red points are PMS systems, while gray points represent all other data points surveyed from the \cite{zari_2018} catalog. The solid line represents the 5 Myr isochrone, the dashed line shows the 10 Myr isochrone, and the dashed-dotted line shows the 20 Myr isochrone. The age, color, and magnitude cuts made by \cite{zari_2018} during the selection process are visible in this plot. }
\label{iso}
\end{figure}

For PMS systems we also obtain age estimates during this analysis. These systems will eventually evolve onto the main sequence, and do so by tracking perpendicular to isochrones, meaning color-magnitude pairs lead to a unique age estimates. This is not valid for UMS systems because they have already reached the main sequence and have been there for an indeterminate amount of time. We follow the same process to obtain ages as we did masses, using the same closest three neighbors in the KDTree and performing a weighted average using the respective distances in the KDTree as weights. Values for the masses, and ages for PMS systems, are provided in Tables  \ref{ums_table} and \ref{pms_table} (supplementary materials), without estimations on uncertainties due to the inherent lack of uncertainty estimates for the isochrones. A plot of color versus $M_G$ for all targets is shown in Figure \ref{iso}. We recover the cutoff values used by \cite{zari_2018}. We note that we do not attempt to account for any effects that a seen or unseen companion would have on these calculations. For binary systems we assume the companions are identical. We note that the only reddening corrections performed on these systems are the ones provided by \cite{zari_2018}, and so these estimates are susceptible to outliers from reddening effects.

\subsection{Distribution of Mass with Age}
\begin{figure}
\centering
\includegraphics[width = .49 \textwidth]{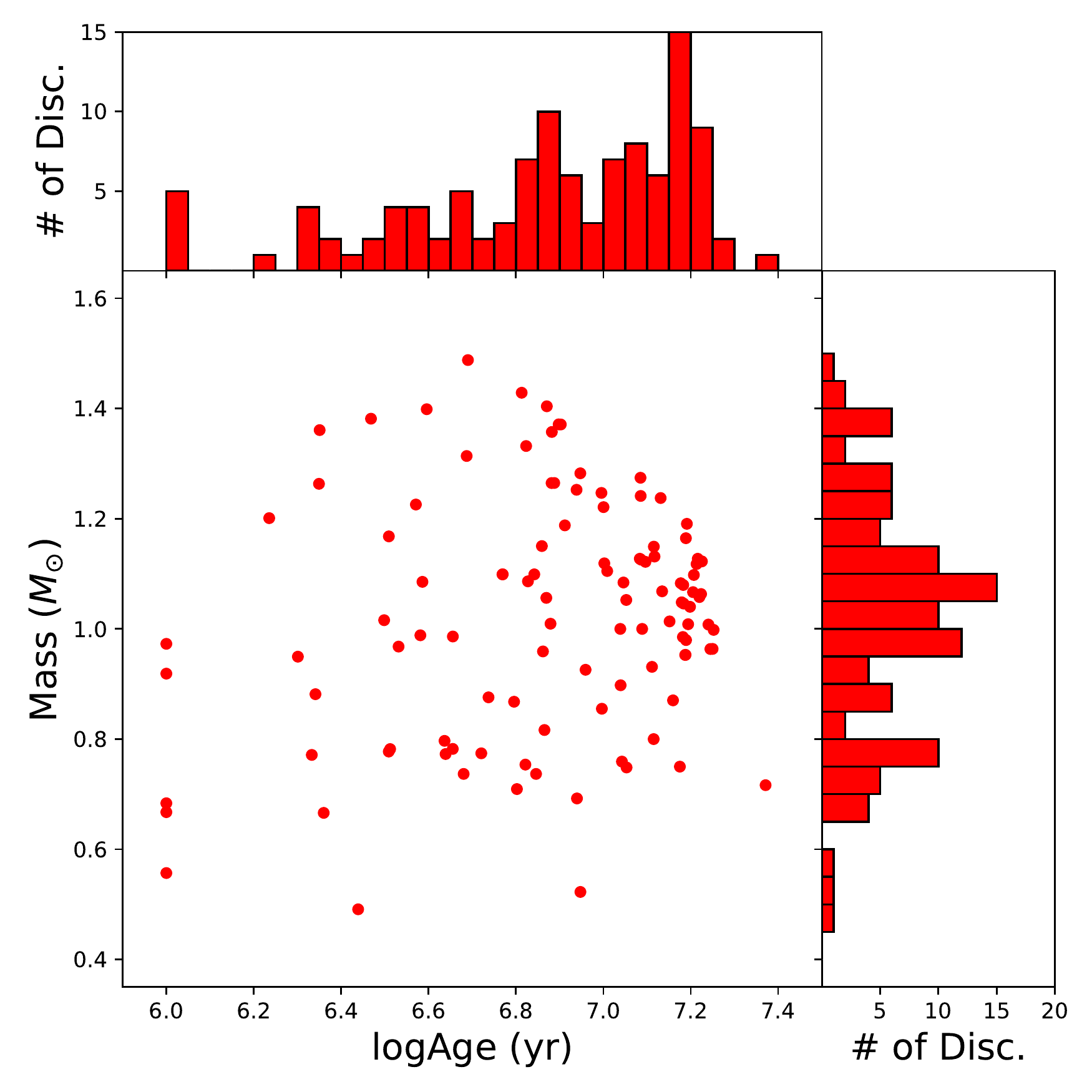}
\caption{A scatter plot showing mass vs age for the PMS non-EBs found in this survey. The histogram on the right side of the plot shows the distribution of masses, with bin sizes of .05 $M_{\odot}$. The histogram on the top of the plot shows the distributions of ages, with bin sizes of .05 in logAge (yr) space. }
\label{var_age_mass}
\end{figure}

Figure \ref{var_age_mass} shows the distribution of mass with age for all non-EB discoveries, with distributions for both of the quantities to the right and top of the graph, respectively. These systems have an average age of ~7.39 Myr and average masses of ~1.02$M_{\odot}$. We also find that the range of ages is from 1 to 23.5 Myr, and the masses range from .49 to 1.48 $M_{\odot}$. We see that five systems seem to have ages of exactly 1 Myr, suggesting these are systems with incorrectly estimated ages. These outliers are likely a result of unresolved binaries, or arise from reddening effects producing an inaccurate age/mass estimation.

\subsection{Light Curve Degeneracies}\label{short_period_variables_sec}
\begin{figure}
\centering
\includegraphics[width = .49 \textwidth]{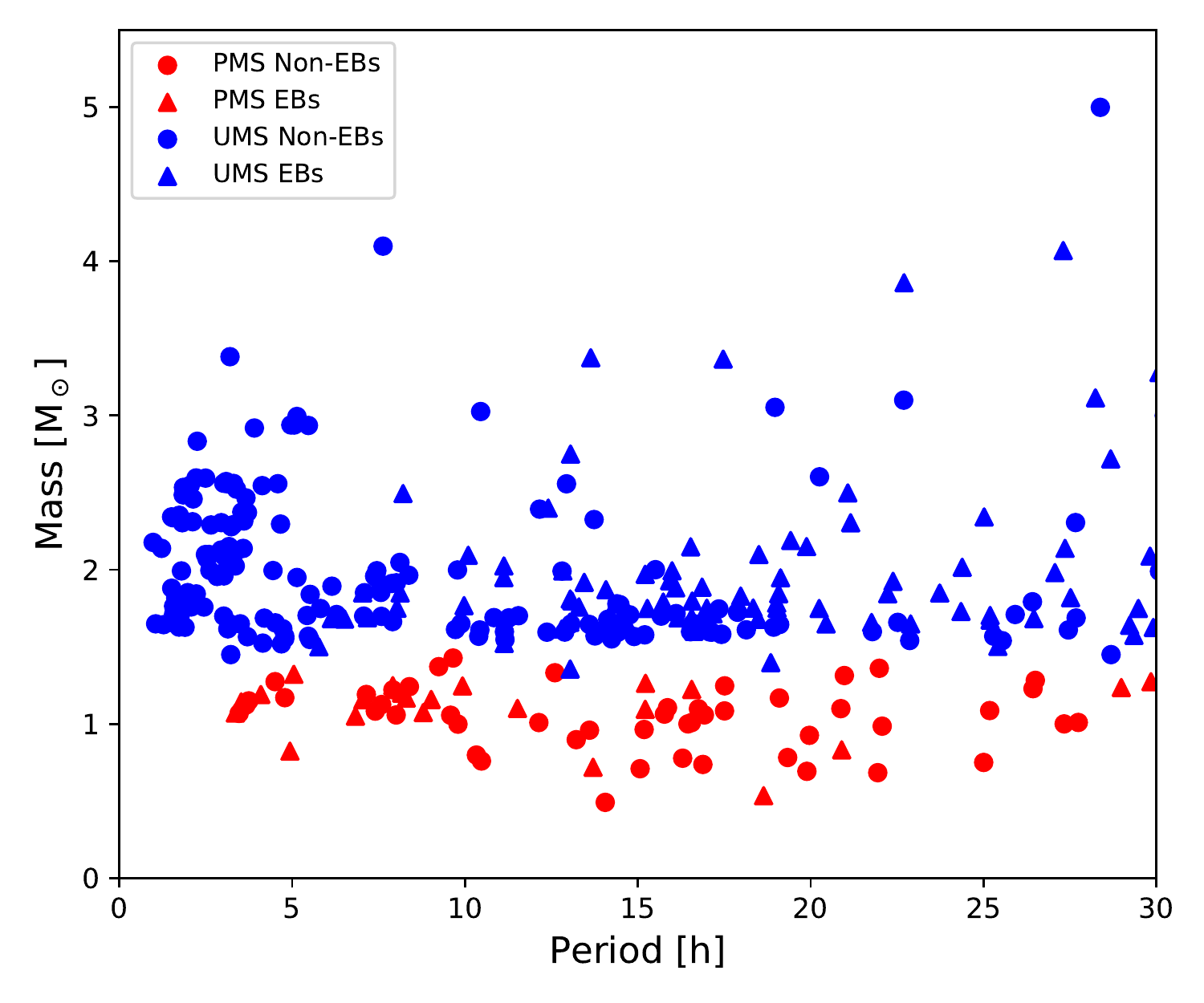}
\caption{The relationship between period and mass for systems with periods less than 30 hr. The circles represent non-EB variables and the triangles are EB systems. Red circles/triangles are systems from the PMS while blue are systems from the UMS.}
\label{mass_period_plots}
\end{figure}

An initially surprising trend can be seen by looking at Figures \ref{discoveries_period_histograms} (top) and \ref{mass_period_plots}. We see a large number of variables with periods on the order of a few hours. Figure \ref{mass_period_plots} shows a plot of mass as a function of period for discoveries with periods less than 30 hr. Going from longer period to the shorter period systems in this figure we notice the number of EBs decreases. For EB systems going to shorter periods means the light curves begin to take on a smoother shape and at some point the light curves of an EB and a non-EB system begin to resemble one another. It is possible that some of the short period EB systems shown in Figure \ref{mass_period_plots} are misclassified due to this similarity. While we attempt to vet these system by looking for characteristics of transits, some of these systems might actually be displaying intrinsic variability that exhibits the characteristics of an EB system, such as flat bottoms in the brightness dips of the light curves. The degeneracy in light curve shapes at these short periods makes further study of these systems necessary in order to be confident of their classifications.

\subsection{Flares}
\begin{figure}
\centering
\includegraphics[width = .49 \textwidth]{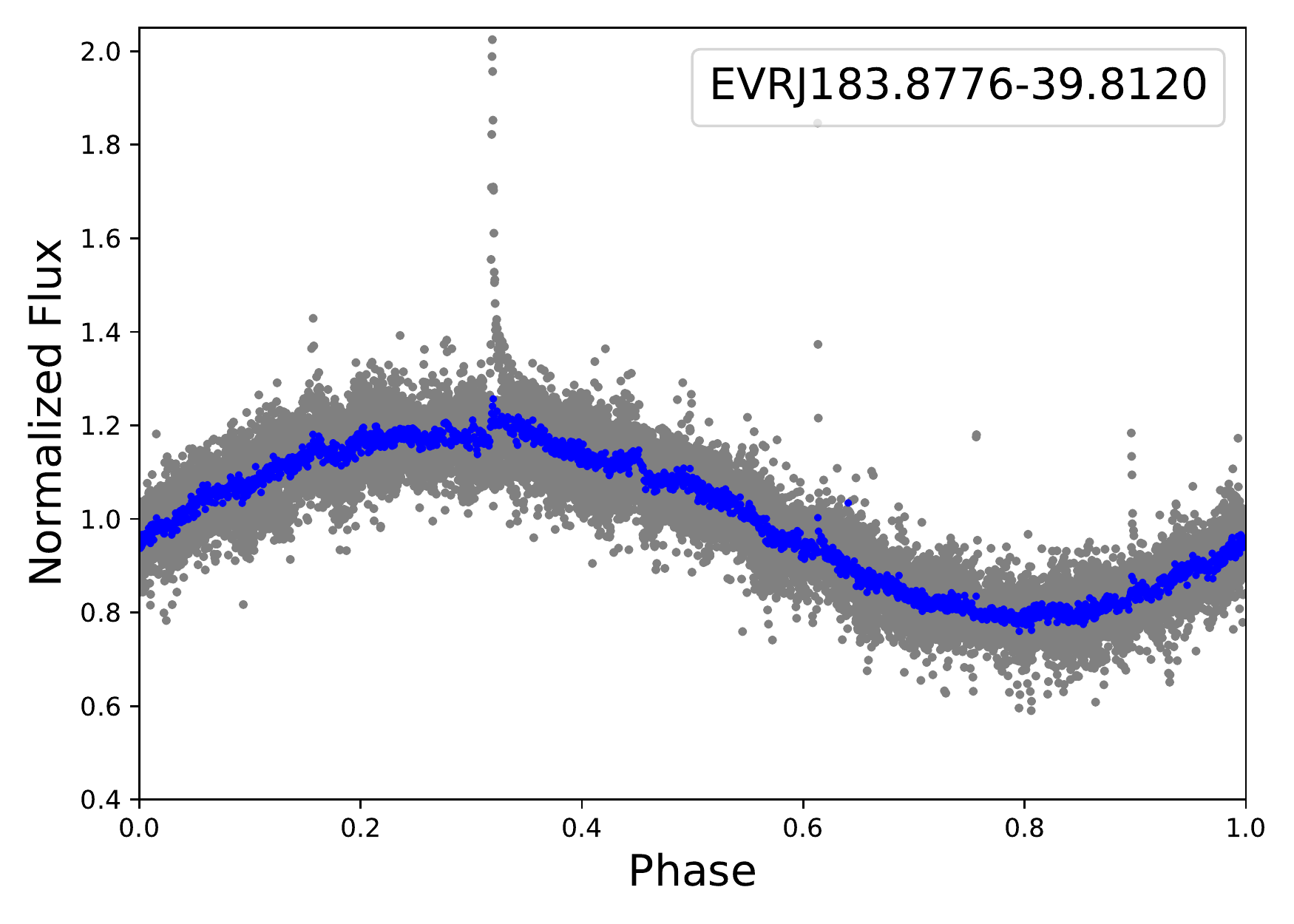}
\includegraphics[width = .49 \textwidth]{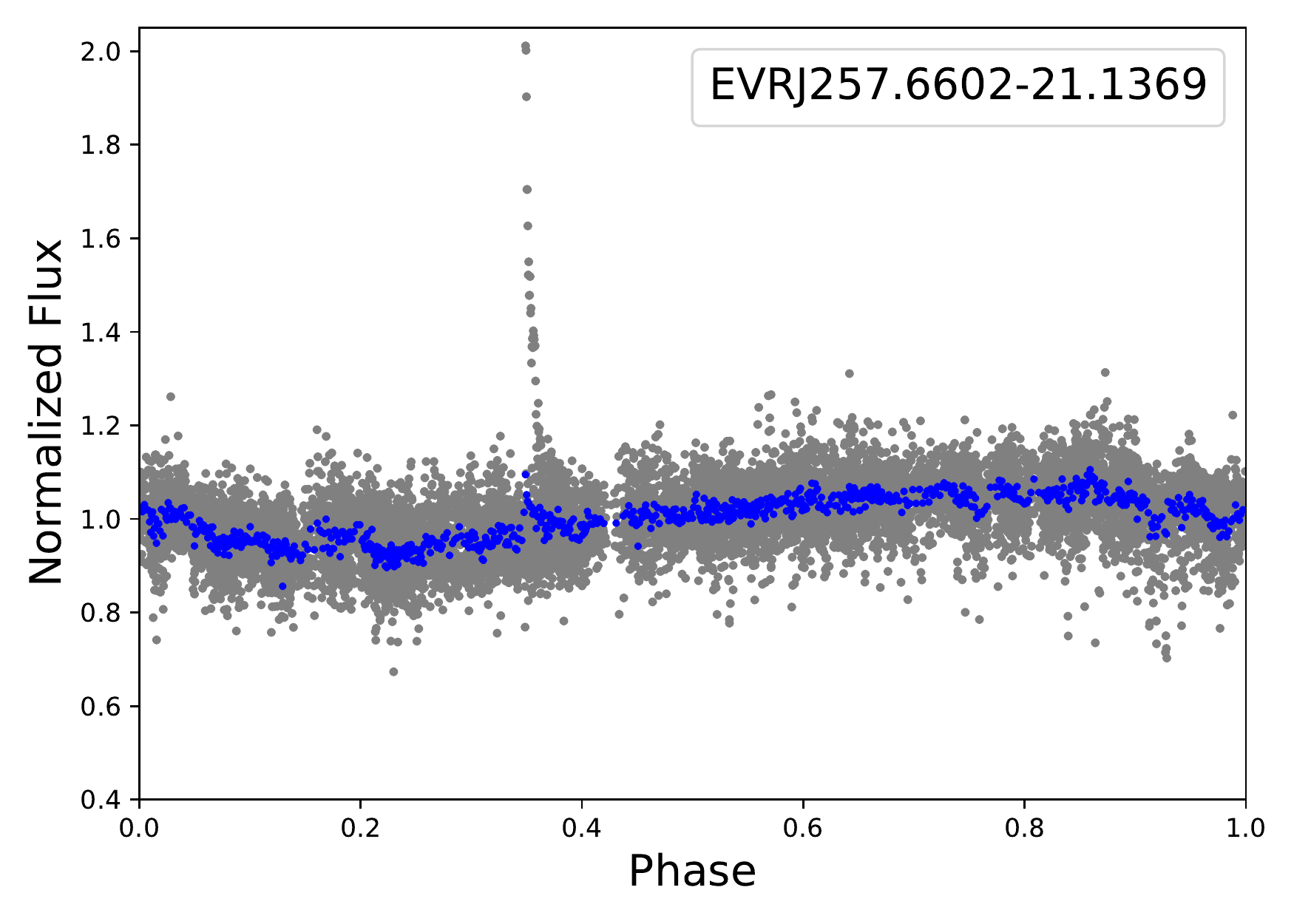}
\caption{Light curves  for \textit{EVRJ183.8776-39.8120} (top) and \textit{EVRJ257.6602-21.1369} (bottom). \textit{EVRJ183.8776-39.8120} shows a strong flaring event at phase 0.32, while \textit{EVRJ257.6602-21.1369} shows a strong flaring event at phase 0.35. The flaring events indicate young, active stars. The gray points show the raw Evryscope data, while the blue points are binned in phase with binning sizes of 15 points per bin. }
\label{flares}
\end{figure}

A strong flaring event was captured in the Evryscope light curve for \textit{EVRJ183.8776-39.8120}, the light curve is shown in Figure \ref{flares} (top). This star has a rotation period of 121.622$\pm$0.005 hr. The mass of this star is estimated to be 0.75$M_{\odot}$ with an age estimation of 11.3 Myr. This light curve shows a rotational variability at the given period, common for young active stars.

A strong flaring event was also captured in the Evryscope light curve for \textit{EVRJ257.6602-21.1369}, the light curve is shown in Figure \ref{flares} (bottom). This star has a rotation period of 87.515$\pm$0.002 hr. The mass of this star is estimated to be 0.77 $M_{\odot}$, with an age estimation of 4.36 Myr. This star also shows a rotational variability along with the flaring event.

The two examples given here show flaring events occurring on stars with rotational periods. Both of these examples are from the PMS, which is expected because young stars that are still undergoing formation processes are most likely to show stellar activity. These two discoveries are not the only stars showing flaring activity in this survey, but they demonstrate the highest amplitude flares.

\section{Summary} \label{section_summary}
This survey has analyzed 44,971 targets in the Evryscope database in search of nearby young variable stars existing on the UMS and PMS. We recovered a total of 615 variable systems, with 313 of those being non-EB variable star candidates and the remaining 302 systems being EB candidates. A large number of new discoveries were made, including 84 young EB candidates and 153 non-EB variables. These discoveries also include 5 new candidate PMS EBs, which are useful systems for understanding PMS stellar evolution. We found five candidate eccentric binary systems, four of which were previously known and one previously unknown (\textit{EVRJ117.9822-19.7374}). These systems exhibit long orbital periods ranging from 299 to 674 hr. We recovered 158 EB systems with periods longer than 50 hr, 9 of which are from the PMS. These systems will be useful in helping constrain the mass/ radius/age relation. For each of our recoveries we have fitted the light curves with either sinusoids or inverted Gaussians, depending on the variability type, and provided the fitting parameters in Tables \ref{ums_table} and \ref{pms_table} (supplementary materials). For these systems we also provide galactic distance to the target, as well as estimations on mass, and age for PMS systems. For PMS non-EBs we have discussed the distribution between these two parameters. All PMS discoveries have masses between 0.4M$_{\odot}$ and 1.6M$_{\odot}$ and ages ranging from 2 to 17 Myr for EBs and 1 to 23 Myr for non-EBs. We discuss that it is possible some of our short period EB discoveries are actually non-EBs, because there exists degeneracies in the light curves of the two types of systems at short periods. Instead, they are possibly rotators or pulsators displaying intrinsic variablilty. We have shown specific examples of stellar activity in our variable recoveries, in the form of flaring events, typical of the young active stars analyzed in this data set.

\section*{Acknowledgements}
The authors thank Andrew Mann for useful discussions which contributed to this manuscript.

The authors acknowledge funding support by the National Science Foundation CAREER grant 1555175, and the Research Corporation Scialog grants 23782 and 23822. HC is supported by the National Science Foundation Graduate Research Fellowship under Grant No. DGE-1144081. The Evryscope was constructed under National Science Foundation/ATI grant AST-1407589.

This research made use of Astropy,\footnote{http://www.astropy.org} a community-developed core Python package for Astronomy (\citealt{astropy:2013}), (\citealt{astropy:2018}).

This survey used the SIMBAD database, operated at CDS, Strasbourg, France.

This survey used the International Variable Star Index (VSX) database, operated at AAVSO, Cambridge, Massachusetts, USA.

This survey used the VizieR catalogue access tool, CDS, Strasbourg, France (DOI : 10.26093/cds/vizier). The original description of the VizieR service was published in AAS 143, 23.


\appendix

\setcounter{table}{0}
\renewcommand{\thetable}{A\arabic{table}}
\section{Discovery Tables}\label{disc_tables}
The two tables in the appendix contain all of the relevant information for the discoveries from this survey. The columns are as follows: ``Type'' indicates whether the target is an eclipsing binary with one eclipse (EB), two eclipses (EB2), a sinusoidally varying star (S), or a periodic non-sinusoidally varying system (NS); ``Period'' contains the recovered period in hours, with associated uncertainties; ``Amp.'' and ``Sec. Amp.'' columns contain the fitted amplitudes for sinusoids and inverted Gaussian for non-EBs and EBs, respectively; ``Mass'' contains the estimated mass for the target in units of stellar mass (for EB systems, no attempt is made to correct for the blending of two stars); ``logAge'' column contains the estimated logarithm of the age for the system, in units of years (only provided for the PMS systems); ``dist.'' column contains the estimated distance to the target in units of pc; ``Prim. Dur.'' and ``Sec. Dur.'' contain the eclipse durations in unit-less values of phase for EB and EB2 types; ``t0'' column contains the zeroth epoch for the primary eclipse; and the ``Ref.'' column indicates whether a citation was found for the target. A ``*'' beside the ID of a target indicates possible contamination between the target star and other stars contained in the same Evryscope pixel. This is generally a small or negligible effect because of the bright stars used in this survey. The full tables are provided as supplementary material.

The entries in this column correspond to references as follows: T94: \citealt{1894cdbp.book.....T}; W27: \citealt{1927BHarO.851...10W}; K30: \citealt{1930BAN.....5..219K}; H34: \citealt{1934AN....253..195H}; J35: \citealt{1935AN....255..417J}; F37: \citealt{1937TrSht...8....5F}; W39: \citealt{1939AnBos...5E...1W}; S39: \citealt{1939cfsc.book.....S}; H41: \citealt{1941KVeBB...7....3H}; C49: \citealt{1949AnHar.112....1C}; G53: \citealt{1953AnHar.113...67G}; J54: \citealt{1954AnCap..17.....J}; J55: \citealt{1955AnCap..18....0J}; O56: \citealt{1956RA......3..313O}; H56: \citealt{1956VeSon...3....1H}; F60: \citealt{1960PZ.....13..166F}; S63: \citealt{1963AJ.....68..428S}; H63: \citealt{1963VeSon...6....1H}; H64: \citealt{1964ApJ...140.1472H}; S65: \citealt{1965IBVS..107....1S}; S67: \citealt{1967IBVS..178....1S}; M68: \citealt{1968AJ.....73..777M}; Z68: \citealt{1968PZ.....16..435Z}; S69: \citealt{1969IBVS..330....1S}; S71: \citealt{1971AJ.....76..338S}; F71: \citealt{1971IBVS..558....1F}; A72: \citealt{1972IBVS..702....1A}; P72: \citealt{1972IBVS..719....1P}; G74: \citealt{1974IBVS..941....1G}; B75: \citealt{1975ApJ...201..653B}; H75: \citealt{1975mcts.book.....H}; W76a: \citealt{1976A&AS...26..227W}; W76b: \citealt{1976AJ.....81.1134W}; H77: \citealt{1977IBVS.1315....1H}; H78: \citealt{1978mcts.book.....H}; L80: \citealt{1980A&AS...41...85L}; M80: \citealt{1980ApJS...44..241M}; O80: \citealt{1980BICDS..19...74O}; B82: \citealt{1982IBVS.2212....1B}; R82: \citealt{1982MNRAS.201...51R}; H82b: \citealt{1982mcts.book.....H}; W85: \citealt{1985A&AS...61..127W}; V85: \citealt{1985cbvm.book.....V}; W86: \citealt{1986ApJ...306..573W}; H86: \citealt{1986IBVS.2877....1H}; K86: \citealt{1986SAAOC..10...27K}; N87: \citealt{1987A&AS...69..371N}; H88: \citealt{1988mcts.book.....H}; F89: \citealt{1989A&AS...80..127F}; L90: \citealt{1990AJ.....99.2019L}; H93: \citealt{1993AJ....105.1511H}; W94: \citealt{1994AJ....107..692W}; H94: \citealt{1994yCat.3133....0H}; C97: \citealt{1997A&A...328..187C}; K97: \citealt{1997A&AS..123..329K}; E97: \citealt{1997ESASP1200.....E}; P98a: \citealt{1998A&A...333..619P}; P98b: \citealt{1998AcA....48...35P}; M99: \citealt{1999ApJ...516L..77M}; M01: \citealt{2001AJ....122.3466M}; P02: \citealt{2002AcA....52..397P}; K02: \citealt{2002MNRAS.331...45K}; O03: \citealt{2003IBVS.5480....1O}; W04: \citealt{2004AJ....127.2436W}; D04: \citealt{2004IBVS.5542....1D}; O04a: \citealt{2004IBVS.5495....1O}; O04b: \citealt{2004IBVS.5570....1O}; O06: \citealt{2006IBVS.5674....1O}; D07: \citealt{2007AJ....133.1470D}; H08: \citealt{2008AJ....136.1067H}; O08: \citealt{2008OEJV...91....1O}; B09: \citealt{2009OEJV...98....1B}; S09: \citealt{2009yCat....102025S}; H10: \citealt{2010A&A...522A..37H}; M11: \citealt{2011A&A...532A..10M}; W11: \citealt{2011MNRAS.416.2477W}; K12: \citealt{2012AcA....62...67K}; R12: \citealt{2012MNRAS.427.2917R}; M14: \citealt{2014MNRAS.437.1681M}; S14a: \citealt{2014A&A...564A..69S}; S14b: \citealt{2014AcA....64..115S}; D14: \citealt{2014ApJS..213....9D}; S14c: \citealt{2014yCat....102023S}; L15: \citealt{2015A&A...578A.136L}; B15: \citealt{2015A&A...581A.138B}; A16: \citealt{2016MNRAS.456.2260A}; O18: \citealt{2018AJ....155...39O}; C18: \citealt{2018ApJS..237...28C}; J18: \citealt{2018MNRAS.477.3145J}; J19a: \citealt{2019MNRAS.485..961J}; J19b: \citealt{2019MNRAS.486.1907J}; J19c: \citealt{2019arXiv190710609J}; 6AS: (ASAS-SN Survey of Variable Stars VI, in prep.)
\LTcapwidth=.9\textwidth

\clearpage
\LongTables
\begin{deluxetable*}{lllllllllll}
EVRJRADec & Type & Period & Amp. & Sec. Amp. & Mass & d & Prim. Dur. & Sec. Dur. & t0 & Ref. \\
 &  & (h) & (mag.) & (mag.) & ( \(\textup{M}_\odot\)) & (pc) & (phase) & (phase) & MJD & \\
\hline
\endhead
EVRJ000.2379-54.7558 & EB2 & 143.379$\pm$0.009 & 0.225$\pm$0.008 & 0.114$\pm$0.008 & 1.5 & 460$\pm$15 & 0.05 & 0.05 & 5.79114E+04 & M01 \\
EVRJ000.9191-14.0853 & S & 9.783$\pm$0.003 & 0.111$\pm$0.006 & --- & 2.0 & 450$\pm$38 & --- & --- & --- & --- \\
EVRJ000.9860-45.2886 & S & 1.73560$\pm$4e-05 & 0.012$\pm$0.003 & --- & 2.4 & 320$\pm$10 & --- & --- & --- & --- \\
EVRJ001.1557-81.3453 & S & 1.54158$\pm$3e-05 & 0.018$\pm$0.003 & --- & 2.3 & 380$\pm$10 & --- & --- & --- & --- \\
EVRJ006.4188-34.8290 & NS & 3.2237$\pm$0.0001 & 0.028$\pm$0.003 & --- & 1.4 & 165$\pm$2 & --- & --- & --- & P02 \\
 &  &  &  & & \textbf{...} &  & & & &  \\
\hline
\\
\caption{UMS variable system recoveries from this survey. The full table is provided as supplementary material. \\ }
\label{ums_table}
\end{deluxetable*}

\LongTables
\begin{deluxetable*}{llllllllllll}
EVRJRADec & Type & Period & Amp. & Sec. Amp. & Mass & logAge &d & Prim. Dur. & Sec. Dur. & t0 & Ref. \\
 &  & (h) & (mag.) & (mag.) & ( \(\textup{M}_\odot\)) & (yr) &(pc) & (phase) & (phase) & MJD & \\
\hline
\endhead
EVRJ022.1481-54.5656 & EB & 9.0233$\pm$0.0002 & 0.137$\pm$0.003 & --- & 1.2 & 7.2 & 460$\pm$13 & 0.22 & --- & 5.77002E+04 & C18 \\
EVRJ022.4998-30.6746 & EB2 & 6.8322$\pm$0.0002 & 0.45$\pm$0.02 & 0.40$\pm$0.01 & 1.1 & 7.2 & 288$\pm$6 & 0.40 & 0.42 & 5.79963E+04 & P02 \\
EVRJ056.2812-10.3059 & S & 3.6665$\pm$0.0002 & 0.048$\pm$0.003 & --- & 1.1 & 7.2 & 233$\pm$4 & --- & --- & --- & J19c \\
EVRJ067.1932+06.8274* & EB & 3.35719$\pm$9e-05 & 0.208$\pm$0.003 & --- & 1.1 & 7.2 & 460$\pm$15 & 0.44 & --- & 5.80753E+04 & W04 \\
EVRJ073.3113-00.3261 & S & 87.0$\pm$0.1 & 0.089$\pm$0.007 & --- & 1.1 & 7.1 & 245$\pm$4 & --- & --- & --- & J19c \\
 &  &  &  & & \textbf{...} &  & & & & &  \\
\hline
\\
\caption{PMS variable system recoveries from this survey. The full table is provided as supplementary material. \\ }
\label{pms_table}
\end{deluxetable*}

\begin{figure*}
\centering
\includegraphics[width =  .9 \textwidth]{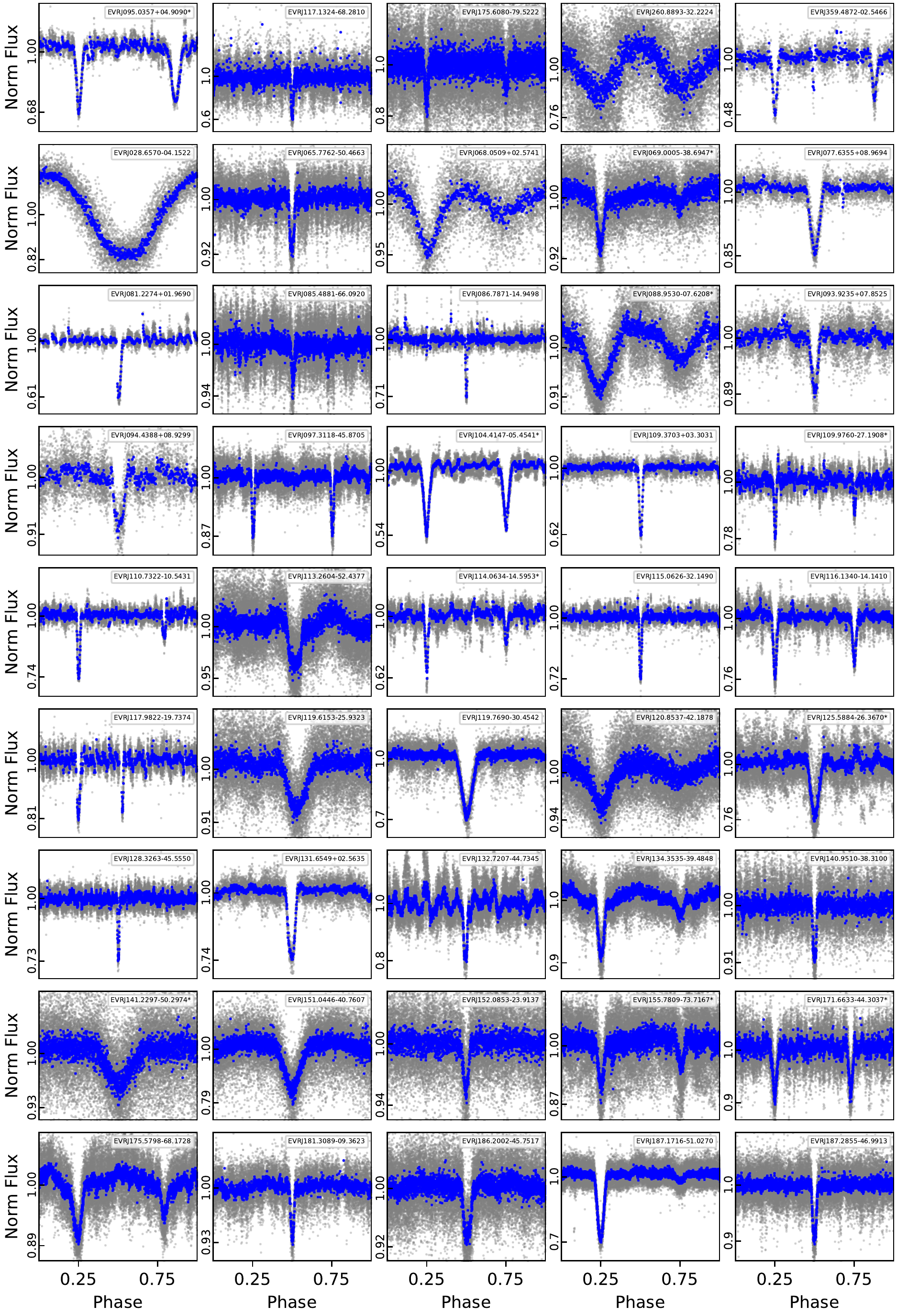}
\caption{Light curves folded at their respective periods for 45 of the new candidate EB discoveries. Raw Evryscope data points are shown in gray and phase-binned points are shown in blue with 15 points per bin. The first row of this plot shows the 5 new candidate PMS EBs, while the remaining light curves show candidates from the UMS. Some of these light curves display poor photometric quality, possibly due to faintness or field crowding.}
\label{large_lc_plt_first_45}
\end{figure*}

\begin{figure*}
\centering
\includegraphics[width =  .9 \textwidth]{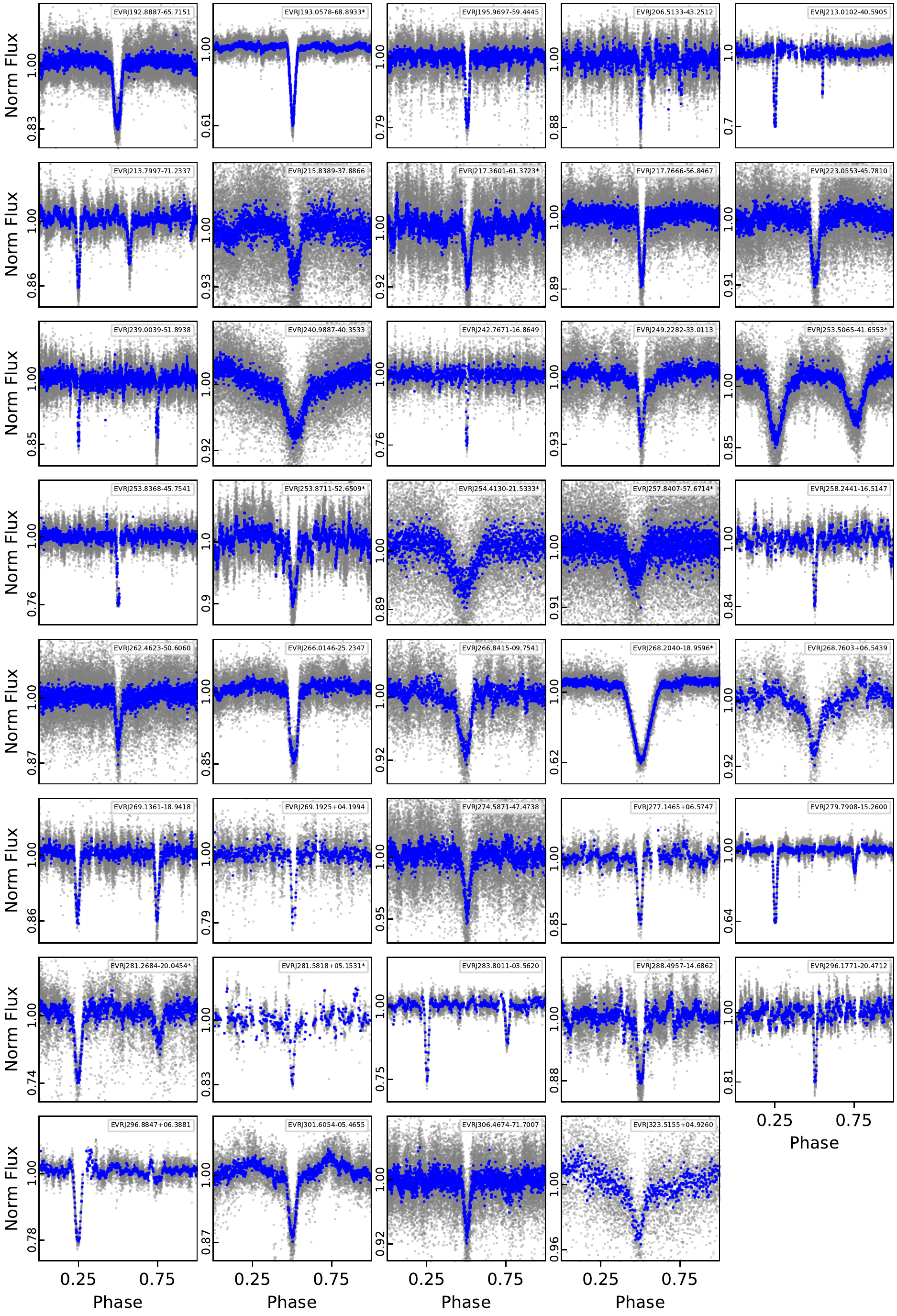}
\caption{Light curves folded at their respective periods for the remaining 39 new candidate EB discoveries. Raw Evryscope data points are shown in gray and phase-binned points are shown in blue with 15 points per bin. Some of these light curves display poor photometric quality, possibly due to faintness or field crowding.}
\label{large_lc_plt_last_39}
\end{figure*}

\clearpage
\bibliographystyle{aasjournal}
\bibliography{bibliography}

\end{document}